\documentclass[sigconf]{acmart}

\usepackage{fancyhdr}
\usepackage[normalem]{ulem}
\usepackage{url}
\usepackage{algorithm}
\usepackage[noend]{algpseudocode}
\usepackage{algorithmicx}
\usepackage{gensymb}
\usepackage{graphicx}
\usepackage[para,online,flushleft]{threeparttable}
\usepackage{tablefootnote} 
\usepackage{multirow} 
\usepackage{mathtools}
\usepackage{subcaption}
\usepackage{xcolor}
\usepackage{xspace}
\usepackage{flushend}
\usepackage{enumitem}
\usepackage{hyperref}
\usepackage{fancyhdr}
\usepackage{tikz}
\usepackage{tikzsymbols}
\usepackage{threeparttable}
\usepackage{slashbox}
\usepackage{datetime}
\usepackage{listings}
\usepackage{setspace}
\usepackage{makecell}
\usepackage{balance}

\newif\ifcameraready
\camerareadyfalse
\newif\ifwebversion
\webversionfalse
\newcommand{\versionnum}[0]{12.7}

\newcommand{\eden}{{EDEN}}

\newcommand{\energy}{energy}
\newcommand{\Energy}{Energy}

\makeatletter
\def\bstctlcite{\@ifnextchar[{\@bstctlcite}{\@bstctlcite[@auxout]}}
\def\@bstctlcite[#1]#2{\@bsphack
 \@for\@citeb:=#2\do{%
    \edef\@citeb{\expandafter\@firstofone\@citeb}%
    \if@filesw\immediate\write\csname #1\endcsname{\string\citation{\@citeb}}\fi}%
 \@esphack}
\makeatother

\copyrightyear{2019}
\acmYear{2019}
\setcopyright{rightsretained}
\acmConference[MICRO-52]{The 52nd Annual IEEE/ACM International Symposium on Microarchitecture}{October 12--16, 2019}{Columbus, OH, USA}
\acmPrice{15.00}
\acmDOI{10.1145/3352460.3358280}
\acmISBN{978-1-4503-6938-1/19/10}

\newcommand\partitle[1]{\vspace{5pt}\noindent\textbf{#1}.}

\newif\ifsubmission
\submissionfalse

\ifsubmission
\newcommand\lois[1]{#1}

\newcommand\agyrm[1]{}
\newcommand\roki[1]{}
\newcommand\rokio[1]{}
\newcommand{\taharm}[1]{}
\newcommand{\taha}[1]{#1}

\newcommand\agycomment[1]{}
\newcommand\loiscomment[1]{}
\newcommand\tahacomment[1]{}
\newcommand\onurcomment[1]{}

\else

\newcommand\lois[1]{\noindent{\color{black}{}{#1}}}

\newcommand\agyrm[1]{}
\newcommand\roki[1]{\noindent{\color{black}{}{#1}}}
\newcommand\rokio[1]{\noindent{\color{orange}{}{#1}}}
\newcommand{\taharm}[1]{{\color{blue}\sout{#1}}\xspace}
\newcommand{\taha}[1]{{\color{black}#1}}

\newcommand\agycomment[1]{}
\newcommand\loiscomment[1]{}
\newcommand\tahacomment[1]{}
\newcommand\onurcomment[1]{}

\definecolor{dkgreen}{rgb}{0,0.6,0}
\definecolor{gray}{rgb}{0.5,0.5,0.5}
\definecolor{mauve}{rgb}{0.58,0,0.82}
\definecolor{chocolate}{rgb}{0.48, 0.25, 0.0}
\makeatletter
\def\BState{\State\hskip-\ALG@thistlm}
\makeatother
\fi

\begin{document}
\bstctlcite{IEEEexample:BSTcontrol}


\title{EDEN: Enabling \Energy{}-Efficient, High-Performance \\ Deep Neural Network Inference Using Approximate DRAM}

\newcommand{\affilETH}[0]{$^1$}
\newcommand{\affilCMU}[0]{$^2$}

\author{%
{Skanda Koppula}%
\quad%
{Lois Orosa}%
\quad%
{A. Giray Ya\u{g}l{\i}k\c{c}{\i}}%
\quad%
}
\author{%
{Roknoddin Azizi}%
\quad%
{Taha Shahroodi}%
\quad%
{Konstantinos Kanellopoulos}%
\quad%
{Onur Mutlu}%
}



\affiliation{
\vspace{0.2cm}
{ETH Z{\"u}rich}%
}

\renewcommand{\shortauthors}{K}

\begin{abstract}


The effectiveness of deep neural networks (DNN) in vision, speech, and language processing has prompted a tremendous demand for \energy{}-efficient high-performance DNN inference systems.
%
Due to the increasing memory intensity of most DNN workloads, main memory can dominate the system's \energy{} consumption and stall time.
%
One effective way to reduce the \energy{} consumption and increase the performance of DNN inference systems is by using approximate memory, which operates with reduced supply voltage and reduced access latency parameters that violate standard specifications. Using approximate memory reduces reliability, leading to higher bit error rates. 
Fortunately, neural networks have an intrinsic capacity to tolerate increased bit errors. This can enable \energy{}-efficient and high-performance neural network inference using approximate DRAM  devices.

Based on this observation, we propose EDEN, the first general framework that reduces DNN \energy{} consumption and DNN evaluation latency by using approximate DRAM devices, while strictly meeting a user-specified target DNN accuracy. EDEN relies on two key ideas: 1) retraining the DNN for a target approximate DRAM device to increase the DNN's error tolerance, and 2) efficient mapping of the error tolerance of each individual DNN data type to a corresponding approximate DRAM partition in a way that meets the {user-specified DNN} accuracy requirements.

We evaluate EDEN on multi-core CPUs, GPUs, and DNN accelerators with error models obtained from real approximate DRAM devices. We show that EDEN's DNN retraining technique reliably improves the error resiliency of the DNN by an order of magnitude. For a target accuracy within 1\% of the original DNN, our results show that EDEN enables {1)} an average DRAM \energy{} reduction of {21\%,} 37\%, 31\%, and 32\%  in CPU, GPU, and two different DNN {accelerator architectures}, respectively, {across a variety of state-of-the-art networks}, and {2) an} average \roki{(maximum)} speedup of 8\% \roki{(17\%)} and 2.7\% \roki{(5.5\%)} in CPU and GPU {architectures}, respectively, when evaluating latency-bound neural networks.

\end{abstract}



\hyphenation{over-para-meter-ization Shah-roodi}
\def\hyph{-\penalty0\hskip0pt\relax}

\maketitle

\fancyhead{}

\newif\ifcameraready
\camerareadytrue

\newif\ifwebversion
\webversiontrue

\fancyhead{}
\ifcameraready
  \ifwebversion
    \fancypagestyle{firststyle}
    {
        \setlength{\footskip}{40pt}
        \fancyfoot[C]{\thepage}
    }
    \thispagestyle{firststyle}
    \pagestyle{firststyle}
  \fi
\else
  \fancyhead[C]{\textcolor{teal}{\emph{Version \versionnum~---~\today, \ampmtime}}}
  \fancypagestyle{firststyle}
  {
    \fancyhead[C]{\textcolor{teal}{\emph{Version \versionnum~---~\today, \ampmtime}}}
    \fancyfoot[C]{\thepage}
  }
  \thispagestyle{firststyle}
  \pagestyle{firststyle}
\fi

\vspace{-1mm}
\section{Introduction}

Deep neural networks (DNNs) \cite{learning, deeplearning} are an effective solution to challenges in computer vision, speech recognition, language translation, drug discovery, robotics, particle physics, and a number of other domains \cite{deeplearning, asr, superresolution, yolo, imagenet2, kws, robotics, robotics2, particlephysics, chemistry, drugdiscovery}. DNNs and their various flavors (convolutional neural networks~\cite{imagenet2}, fully-connected neural networks \cite{fcn}, and recurrent neural networks \cite{asr}) are commonly evaluated in settings with edge devices that demand low \energy{} and real-time responses \cite{mobilenetv2, pim1}. Unfortunately, DNNs have high computational and memory demands that make these \energy{} and performance requirements difficult to fulfill. As such, neural networks have been the subject of many recent accelerators and DNN-focused architectures. Recent works (e.g., \cite{eyeriss, cnvlutin, eie, cambriconx, diannao, origami, stripes, scope, scnn, yodnn, cbrain, angel, isaac, fused, escher, pim3}) focus on building specialized architectures for efficient computation scheduling and dataflow to execute DNNs~\cite{flexflow, flexflow2}.

Improvements to accelerator efficiency \cite{eyerissv2, flexflow3}, DNN-optimized GPU kernels \cite{cudnn, tvm}, and libraries designed to efficiently leverage instruction set extensions \cite{openvino, tvm} have improved the computational efficiency of DNN evaluation. However, improving the memory efficiency of DNN evaluation is an on-going challenge \cite{quant1, eie, pim1, pim2, pim3, pim4}. The memory intensity of DNN inference is increasing, and the sizes of state-of-art DNNs have grown dramatically in recent years. The winning model of the 2017 ILSVRC image recognition challenge~\cite{ilsvrc}, ResNeXt, contains 837M FP32 parameters (3.3 GB) \cite{resnext}. This is 13.5x the parameter count of AlexNet, the winning model in 2012 \cite{imagenet2}. More recent models have broken the one billion FP32 parameter mark (3.7 GB) \cite{bignetwork}. As the machine learning community trends towards larger, more expressive neural networks, we expect off-chip memory problems to bottleneck DNN evaluation.

The focus of this work is to alleviate two main issues (\energy{} and latency) of off-chip DRAM for neural network workloads.
First, DRAM has high \energy{} consumption. Prior works on DNN accelerators report that between 30 to 80\% of system \energy{} is consumed by DRAM~\cite{eyeriss, smartshuttle, memorycentric, mainmemorychallenges}.
Second, DRAM has high latency. A load or store that misses the last level cache (LLC) can take 100x longer time to service compared to an L1 cache hit~\cite{dramlatencycost, chou2004microarchitecture, dundas1997improving, mutlu2005onreusing, sprangle2002increasing, dundas1999improving,mutlu2003runahead,Mutlu:2005}. Prior work in accelerator design has targeted DRAM latency as a challenge for sparse and irregular DNN inference~ \cite{quest}.

To overcome both DRAM \energy{} and latency issues, recent works use three main approaches. 
First, some works reduce numeric bitwidth, reuse model weights, and use other algorithmic strategies to reduce the memory requirements of the DNN workload~\cite{reversiblenn, nullanet, sublinearmemorycost, ubernet, mobilenetv2, squeezenet, quantization_scheme, quantization2, eccenergy, energyawarepruning, amc, netadapt,ttq,deepcompression}. 
{Second}, other works propose new DRAM designs that offer \pagebreak 
lower \energy{} and latency than commodity DRAM~\cite{aldram, tldram, wang2018reducing, voltron, diva, flydram, chang2016lisa, hassan2019crow}.  
Third, some works propose processing-in-memory approaches that can reduce data movement and access data with lower latency and \energy{}~\cite{pim1, pim2, pim3, pim4, pim5, isaac, tetris, dracc,li2017drisa,seshadri2017ambit}. 
In this work, we propose an approach that is orthogonal to these existing works: we customize the major operational parameters (e.g., voltage, latency) of \emph{existing} DRAM chips to better suit the intrinsic characteristics of a DNN. Our approach is based on two key insights:
\begin{enumerate}
    \item DNNs demonstrate remarkable robustness to errors introduced in input, weight, and output data types. This error tolerance allows accurate DNN evaluation on unreliable hardware \lois{if} the DNN error tolerance is accurately characterized and bit error rates are appropriately controlled.
    \item DRAM manufacturers trade performance for reliability. Prior works show that reducing DRAM supply voltage and timing parameters improves the DRAM \energy{} consumption and latency, respectively, at the cost of reduced reliability, i.e., increased bit error rate~\cite{voltron, aldram, goossens, vampire, flydram}.
\end{enumerate}

To exploit these two insights, we propose EDEN\footnote{\textbf{E}nergy-Efficient \textbf{De}ep \textbf{N}eural Network Inference Using Approximate DRAM }: the first framework that improves energy efficiency and performance for DNN inference by using approximate DRAM, which operates with reduced DRAM parameters (e.g., voltage and latency). EDEN strictly meets a user-specified target DNN accuracy by providing a general framework that 1) uses a new retraining mechanism to improve the accuracy of a DNN {when executed on} approximate DRAM, and 2) maps the DNN to the approximate DRAM using {information obtained from rigorous characterizations of the} DNN {error tolerance} and DRAM {error properties}.

EDEN is based on three key steps. 
First, EDEN \emph{improves the error tolerance} of the target DNN by retraining {the DNN} using the {error characteristics of the} approximate DRAM module. 
Second, EDEN \emph{profiles} the improved DNN to identify the error tolerance levels of all DNN data (e.g., different layer weights of the DNN). 
Third, EDEN \emph{maps} different DNN data to different DRAM partitions that best fit each datum's characteristics, and accordingly selects the voltage and latency parameters to operate each DRAM partition. By applying these three steps, EDEN can map an arbitrary DNN workload to an arbitrary approximate DRAM module to evaluate a DNN with low \energy{}, high performance, \emph{and} high accuracy.

To show example benefits of our approach, we use EDEN to run inference on DNNs using approximate DRAM with 1) reduced DRAM supply voltage ($V_{DD}$) to decrease DRAM \energy{} consumption, and 2) reduced DRAM latency to reduce the execution time of latency-bound DNNs. EDEN adjusts the DRAM supply voltage and DRAM latency through interaction with the memory controller firmware. 
For a target accuracy within 1\% of the original DNN, our results show that EDEN enables 1) an average DRAM \energy{} reduction of 32\% across CPU, GPU and DNN accelerator (e.g., Tensor Processing Unit~\cite{tpu}) architectures, and 2) cycle reductions of up to 17\% when evaluating latency-bound neural networks.

Our evaluation indicates that the larger benefits of EDEN would stem from its capacity to run on most hardware platforms in use today for neural network inference, including CPUs, GPUs, FPGAs, and DNN accelerators. Because EDEN is a general approach, its principles can be applied 1) on any platform that uses DRAM, and 2) across memory technologies that can trade-off different parameters (e.g., voltage, latency) at the expense of reliability. Although our evaluation examines supply voltage and access latency reductions, the EDEN framework can be used also to improve performance and \energy{} in other ways: for example, EDEN could increase the effective memory bandwidth by increasing the data bus frequency at the expense of reliability.

This paper makes the following five key contributions:
\begin{itemize}[noitemsep,topsep=2pt]
\item We introduce EDEN, the first general framework that increases the energy efficiency and performance of DNN inference by using approximate DRAM that operates with reduced voltage and latency parameters at the expense of reliability. EDEN provides a systematic way to scale main memory parameters (e.g., supply voltage and latencies) while achieving a user-specified DNN accuracy target.

\item We introduce a methodology to retain DNN accuracy in the presence of approximate DRAM. Our evaluation shows that EDEN increases the bit error tolerance of a DNN by 5-10x (depending on the network) through a customized retraining procedure called \emph{curricular retraining}.

\item We provide a systematic, empirical characterization of the resiliency of state-of-art DNN workloads \cite{densenet,squeezenet,mobilenetv2} to the errors introduced by approximate DRAM. We examine error resiliency across different numeric precisions, pruning levels, and data types (e.g. DNN layer weights). We find that 1) lower precision levels and DNN data closer to the first and last layers exhibit lower error resiliency, and 2) magnitude-based pruning does not have a significant impact on error resiliency.

\item We propose four error models to represent the common error patterns that an approximate DRAM device exhibits. To do so, we characterize the bit flip distributions that are caused by reduced voltage and latency parameters on eight real DDR4 DRAM modules.

\item We evaluate EDEN on multi-core CPUs, GPUs, and DNN accelerators. For a target accuracy within 1\% of the original DNN, our results show that
EDEN enables {1)} an average DRAM \energy{} reduction of {21\%,} 37\%, 31\%, and 32\%  in CPU, GPU, and two different DNN {accelerator architectures}, respectively, {across a variety of state-of-the-art networks}, and {2) an} average \roki{(maximum)} speedup of 8\% \roki{(17\%)} and 2.7\% \roki{(5.5\%)} in CPU and GPU {architectures}, respectively, when evaluating latency-bound neural networks.
For a target accuracy the same as the original, \eden{} enables {16\% average} energy savings and {4\%} average speedup in CPU {architectures}.
\end{itemize}

\section{Background}
\label{sec:background}

\subsection{Deep Neural Networks}
\label{sec:dnn_background}
\begin{figure}[h]
\includegraphics[width=0.48\textwidth]{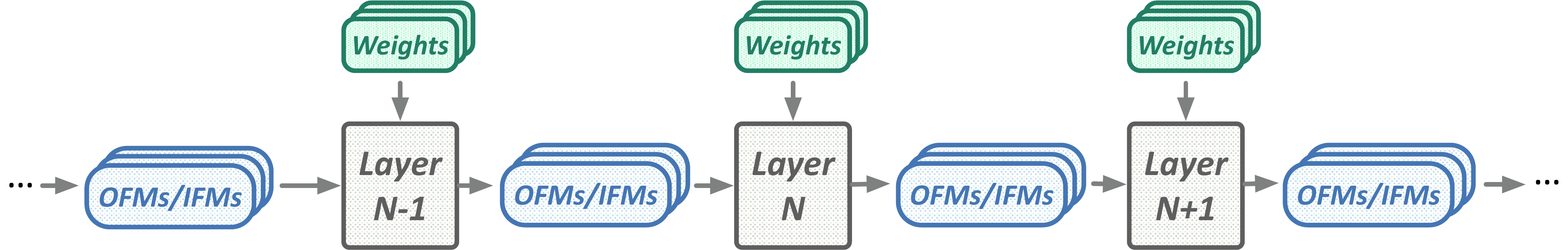}
\caption{Example of three DNN layers. Each layer is composed of its weights, input feature maps (IFMs), and output feature maps (OFMs).}
\label{fig:dnn}
\vspace*{-0.6cm}
\end{figure}

A deep neural network (DNN) is a neural network with more than two layers~\cite{deeplearning}. DNNs are composed of a variety of different layers, including convolutional layers, fully-connected layers, and pooling layers. Figure~\ref{fig:dnn} shows the three main data types of a DNN layer, and how three DNN layers are connected with each other. Each of these layers is defined by a weight matrix learned via a one-time training process that is executed before the DNN is ready for inference.
The three DNN data types that require loads and stores from main memory include each layer's input feature maps (IFMs), output feature maps (OFMs), and the weights. Each layer processes its IFMs using the layer's weights, and produces OFMs. The OFMs of a layer are fed to the next layer as the next layer's IFMs. In this work, we explore the introduction of bit errors into the three data types of each layer. 

Modern DNNs contain hundreds of layers, providing the DNN with a large number of trainable weights. The existence of such a large number of weights is commonly referred to as \emph{overparameterization}, and is, in part, the source of a DNN's accuracy~\cite{poweroverparameterization}. Overparameterization allows the model to have sufficient learning capacity so that the network can approximate complex input-output functions, and adequately capture high-level semantics (e.g., the characteristics of a cat in an input image)~\cite{roleoverparameterization}. Importantly, overparameterization allows the network to obtain some level of error resilience, generalize across different inputs, and be robust to insignificant changes to the input (e.g., background pixels in an image)~\cite{generalization}. Common training-time techniques such as \emph{adding white noise} and \emph{input feature map dropout} try to force the network to \emph{not} rely on any single OFM element and enable robustness in the presence of statistical variance in the IFMs~\cite{dropout}. In this work, we show that we can also adapt DNNs and their training procedure to achieve partial error robustness against bit errors caused by approximate DRAM, by fundamentally taking advantage of the overparameterization in the DNN.

\partitle{Quantization} Quantizing floating-point weights and OFMs into low-precision fixed-point numbers can greatly improve performance and energy consumption of DNNs~\cite{quantizedNN}. Many prior works demonstrate that it is possible to quantize DNNs to limited numeric precision (e.g., eight-bit integers) without significantly affecting DNN accuracy~\cite{quantization_scheme,quantizedNN,binarizedNN,wu2016quantized,ttq,deepcompression,quantization2,quest}. In our evaluations, we quantize all DNN models to four different numeric precisions : int4 (4-bit), int8 (8-bit), int16 (16-bit), and FP32 (32-bit).

\partitle{Pruning} Pruning~\cite{braindamage} reduces the memory footprint of a DNN by sparsifying the weights and feature maps. This is done by zeroing the lowest magnitude weights and retraining~\cite{li2016pruning,scalpel,deepcompression}. We study the effects of pruning in our evaluations.

\partitle{Training} Training is the process of estimating the best set of weights that maximize the accuracy of DNN inference. Training is usually performed with an iterative gradient descent algorithm~{\cite{robbins1951stochastic}} using a particular training dataset. The training dataset is divided into batches. One iteration is the number of batches needed to complete one epoch. One epoch completes when the entire dataset is passed once through the training algorithm.

\subsection{DRAM Organization and Operation}

\begin{figure}[t]
\includegraphics[width=0.48\textwidth]{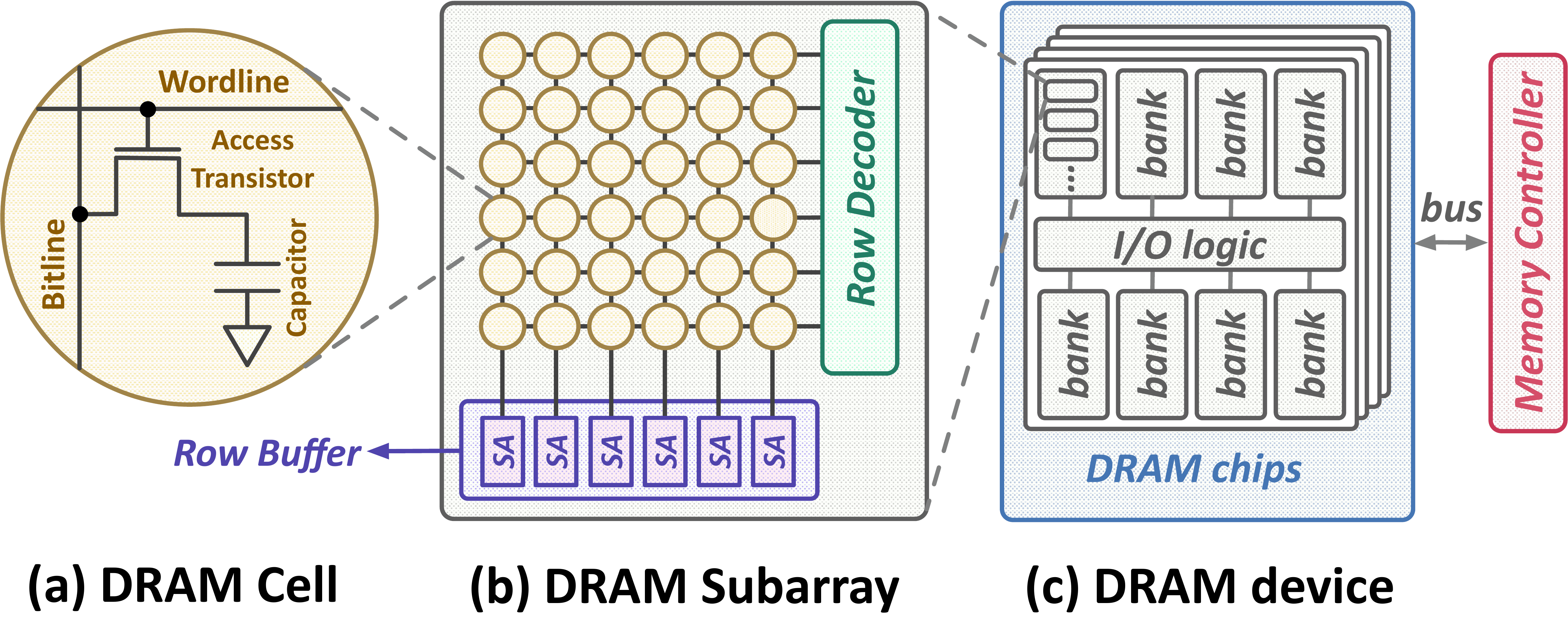}
\vspace*{-0.7cm}
\caption{DRAM organization.}
\vspace{-0.5cm}
\label{fig:dram_structure}
\end{figure}

\partitle{DRAM Organization} A DRAM device is organized hierarchically. Figure~\ref{fig:dram_structure}a shows a \emph{DRAM cell} that consists of a \emph{capacitor} and an \emph{access transistor}. A capacitor encodes a bit value with its charge level. The DRAM cell capacitor is connected to a \emph{bitline} via an access transistor that is controlled by a \emph{wordline}.
Figure~\ref{fig:dram_structure}b shows how the DRAM cells are organized in the form of a 2D array (i.e., a \emph{subarray}). Cells in a column of a subarray share a single bitline. Turning on an access transistor causes charge sharing between the capacitor and the bitline, which shifts the bitline voltage up or down based on the charge level of the cell's capacitor. Each bitline is connected to a \emph{sense amplifier} (SA) circuit that detects this shift and amplifies it to a full 0 or 1.
The cells that share the same wordline in a subarray are referred to as a DRAM \emph{row}. A \emph{row decoder} drives a \emph{wordline} to enable all cells in a DRAM row. Therefore, charge sharing and sense amplification operate at row granularity. The array of sense amplifiers in a subarray is referred to as \emph{row buffer}. Each subarray typically consists of 512-1024 rows each of which is typically as large as 2-8KB.

Figure~\ref{fig:dram_structure}c shows the organization of subarrays, {banks,} and chips that form a \emph{DRAM device}. Each bank partially decodes a given row address and selects the corresponding subarray's \emph{row buffer}.
On a read operation, the I/O logic sends the requested portion of the target row from the corresponding subarray's row buffer to the memory controller.
A \emph{DRAM chip} contains multiple banks that can operate in parallel. A DRAM device is composed of multiple DRAM chips that share the same command/address bus and are simultaneously accessed to provide high bandwidth and capacity. 
In a typical system, each memory controller interfaces with a single DRAM bus.
We refer the readers to \cite{salp, raidr, tldram, drambook, blp, chargecache, hassan2019crow,seshadri2019dram,wang2018reducing,flydram,changHPCA2014,chang2016lisa,voltron,aldram,diva,refresh1,halfdram} for more detail on DRAM structure and design.

\begin{figure}[h]
\vspace*{-0.2cm}
\includegraphics[width=0.48\textwidth]{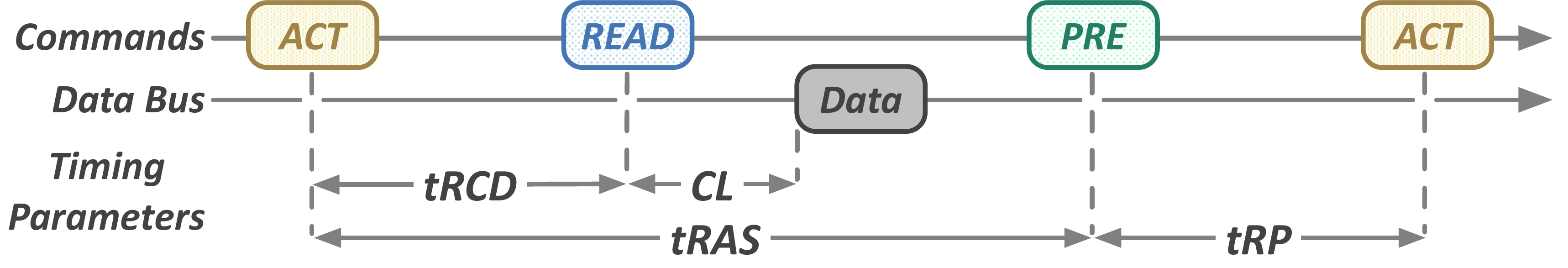}
\vspace*{-0.7cm}
\caption{DRAM read timing. We explore reductions of $t_{RCD}$, $t_{RAS}$, and  $t_{RP}$ as part of EDEN's evaluation. $CL$ is a characteristic of the device, and not adjustable in the memory controller~\cite{jedecddr4}.}
\vspace*{-0.5cm}
\label{fig:dram_timing}
\end{figure}

\partitle{DRAM Operation} Accessing data stored in each row follows the sequence of memory controller commands illustrated in Figure \ref{fig:dram_timing}. First, the activation command (ACT) activates the row by pulling up the wordline and enabling sense amplification. After a manufacturer-specified $t_{RCD}$ nanoseconds, the data is reliably sensed and amplified in the \emph{row buffer}. Second, the read command (READ) reads the data from the row buffer to the IO circuitry. After a manufacturer-specified {$CL$ nanoseconds,} the data is available on the memory bus. Third, the precharge command (PRE) prepares the DRAM bank for activation of another row. A precharge command can be issued a manufacturer-specified $t_{RAS}$ nanoseconds after an activation command, and an activation command can be issued $t_{RP}$ nanoseconds after a precharge command. $t_{RCD}$, $t_{RAS}$, $t_{RP}$, and $CL$ are examples of DRAM timing parameters and their nominal values provided in DRAM DDR4 datasheets are 12.5ns, 32ns, 12.5ns, and 12.5ns respectively~\cite{jedecddr4}.

\subsection{Reducing DRAM Parameters}
\label{sec:background_reducing}
We build on a large body of work on characterizing DRAM behavior in sub-reliable operation regimes of \emph{supply voltage} and \emph{latency} parameters \cite{flydram, voltron, diva, solardram, chang2017thesis, dsn, puf, drange, goossens, softmc, aldram, refresh1}.

\partitle{DRAM Voltage Reduction} Voltage reduction is critical to reducing DRAM power consumption since power is proportional to the square of supply voltage (i.e., ${{V_{DD}}^{2}}\times{f}$)~\cite{microntn4007, vogelsang2010understanding}. Prior research~\cite{david2011memory, voltron} shows that reducing voltage increases the propagation delay of signals, which can cause errors when using unmodified timing parameters. One work avoids these errors by increasing the $t_{RCD}$ and $t_{RP}$ latencies~\cite{voltron} to ensure  {\emph{reliable}} operation.
In contrast, our goal in this work is to aggressively reduce power consumption and latency by decreasing both supply voltage and timing parameters, which inevitably causes errors in the form of bit flips in the weakest cells of DRAM, making DRAM approximate. Resulting error patterns often exhibit locality. Chang et al.~\cite{voltron} observe that these bit flips accumulate in certain regions (e.g., banks and rows) of DRAM. 

\partitle{DRAM Access Latency Reduction} Latency reduction is critical to increase system performance, as heavily emphasized by a recent study on workload-DRAM interactions~\cite{workloads-dram}. Previous works characterize real DRAM devices to find the minimum reliable \emph{row activation ($t_{RCD}$)} and \emph{precharge ($t_{RP}$)} latency values \cite{aldram, flydram, diva, solardram,chang2017thesis}. According to these studies, the minimum DRAM latency values are significantly smaller than the values that datasheets report, due to conservative guardbands introduced by DRAM manufacturers.
Further reducing these latency values cause bit flips in weak {or unstable} DRAM cells.

\partitle{DRAM Refresh Rate Reduction} Other than voltage and latency, previous research also shows that reducing the refresh rate of DRAM chips both can increase performance and reduce energy consumption at the cost of introducing errors \cite{refresh1, refresh2, refresh3, raidr, samira-sigmetrics,parbor,reaper,softmc,dsn} that are tolerable by many workloads that can tolerate bit errors~\cite{approxnnrefresh, approx3, approx4, approx5}.

\section{EDEN Framework}

To efficiently solve the \energy{} and latency issues \lois{of} off-chip DRAM for neural network workloads, we propose \eden{}. \eden{} is the first general framework that improves \energy{} efficiency and performance for neural network inference by using approximate DRAM. \eden{} is based on two main insights: 1) neural networks are tolerant to errors, and 2) DRAM timing parameters and voltage can be reduced at the cost of introducing more bit errors.

We first provide an overview of \eden{} in Section~\ref{sec:eden}, and explain \eden{}'s three steps in Sections~\ref{sec:boosting}, \ref{sec:networkcharacterization}, and \ref{sec:netmapping}. Finally, Section~\ref{sec:inference} explains the changes required by the target DNN inference system to support a DNN generated by \eden{}.

\subsection{EDEN: A High Level Overview}
\label{sec:eden}

\begin{figure}[h]
\centering
\includegraphics[width=1\linewidth]{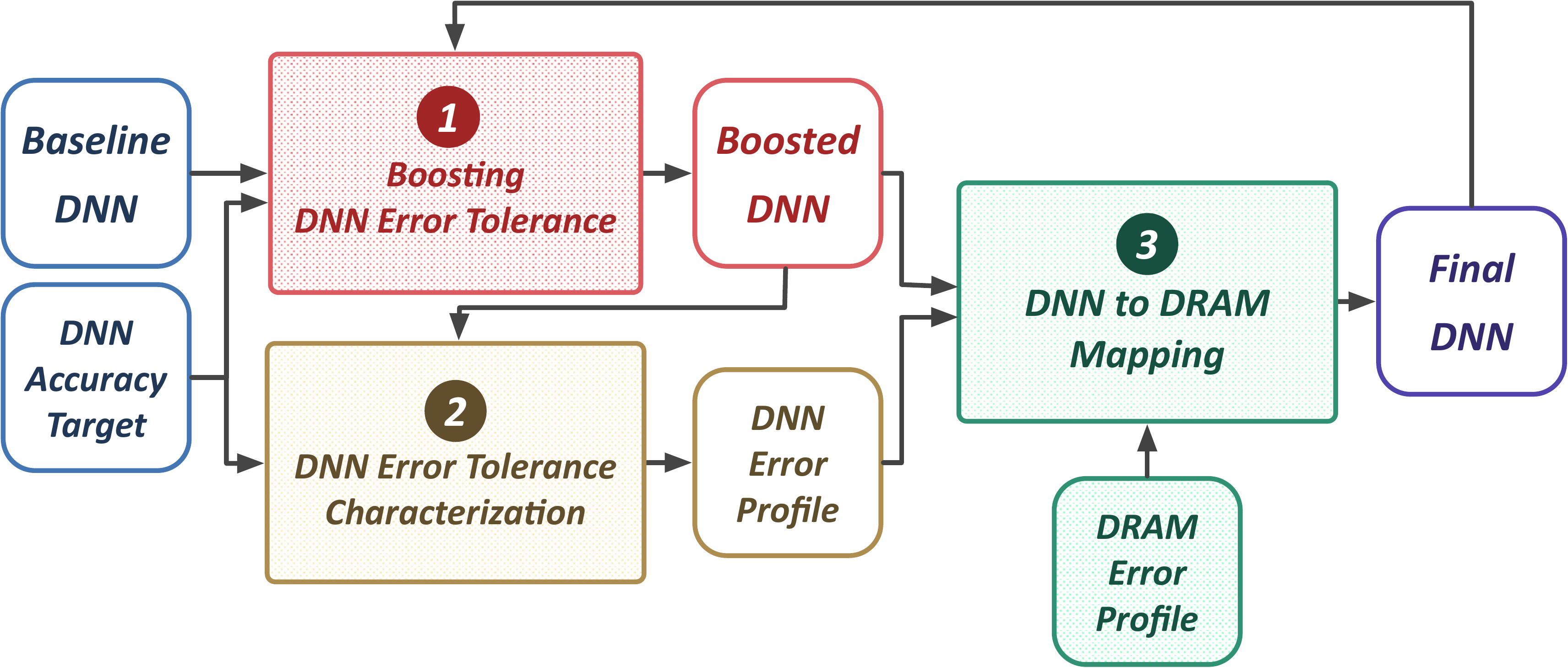}
\vspace{-0.7cm}
\caption{Overview of the EDEN framework.}
\vspace{-0.6cm}
\label{fig:eden_overview}
\end{figure}

EDEN enables the effective \lois{execution} of DNN workloads \lois{using} approximate DRAM through three key steps:
1)~boosting DNN error tolerance,
2)~DNN error tolerance characterization, and
3)~DNN-DRAM mapping.
These steps are repeated iteratively until \eden{} finds the most aggressive DNN and DRAM configuration that meets the target accuracy requirements.
EDEN transforms a DNN that is trained on reliable hardware into a device-tuned DNN that is able to run on a system that uses approximate DRAM at a target accuracy level. EDEN allows tight control of the trade-off between accuracy and performance by enabling the user/system to specify the maximal tolerable accuracy degradation.  Figure~\ref{fig:eden_overview} provides an overview of the three steps of EDEN, which we describe next.

\partitle{1. Boosting DNN Error Tolerance}
\eden{} introduces \emph{curricular retraining},  a new retraining mechanism that boosts a DNN's error tolerance for a target approximate DRAM module. Our curricular retraining mechanism uses the error characteristics of the target approximate DRAM to inject errors into the DNN training procedure and boost the DNN accuracy. The key novelty of curricular retraining is to inject errors at a progressive rate during the training process with the goal of increasing DNN error tolerance while avoiding accuracy collapse with error correction. \eden{} boosts the intrinsic bit error tolerance of the baseline DNN by 5-10x. We describe our boosting mechanism in Section~\ref{sec:boosting}.

\partitle{2. DNN Error Tolerance Characterization}
\eden{} characterizes the error resilience of each \emph{boosted DNN data type} (i.e., IFMs, OFMs, and DNN weights) to identify the limits of bit error tolerance. \eden{} measures the effect of bit errors on overall accuracy using the DNN validation dataset. We describe error tolerance characterization in Section~\ref{sec:networkcharacterization}.

\partitle{3. DNN to DRAM Mapping} 
\eden{} maps the error tolerance of each DNN data type to a corresponding approximate DRAM partition (e.g., chip, bank, or subarray) in a way that meets the specified accuracy requirements, while maximizing performance. We describe DNN to DRAM mapping in Section~\ref{sec:netmapping}.

Together, the three steps of EDEN enable a baseline DNN to become a specialized DNN that is error-tolerant and device-tuned to a target approximate DRAM. EDEN enables \energy{} efficient, high-performance DNN inference on the target approximate DRAM with a user-defined accuracy.

\subsection{Boosting DNN Error Tolerance}
\label{sec:boosting}

According to our evaluations, the error tolerance of common DNNs is not sufficient to enable significant DRAM voltage and timing parameter reductions.
To overcome this issue, we propose \emph{curricular retraining}, a new retraining mechanism that improves the error tolerance of a DNN when {running} with approximate DRAM that injects errors into memory locations accessed by {the} DNN.

The key idea of curricular retraining is based on the observation that introducing high error rates immediately at the beginning of retraining process occasionally causes training divergence and a phenomenon called  \emph{accuracy collapse}. To mitigate this problem, \emph{curricular retraining}  slowly increases the error rate of the approximate DRAM from 0 to a target value in a step-wise fashion. In our experiments, we observe a good training convergence rate when we increase the error rate every two epochs (i.e., two passes of the entire training dataset). \eden{} uses approximate DRAM in the forward pass, and it uses reliable DRAM for the backward pass.

We demonstrate in Section~\ref{sec:softmcdemonstration} that our curricular retraining mechanism is effective at improving the accuracy of DNN inference executed on systems with approximate DRAM.

Our experiments show that curricular retraining does not help to improve DNN accuracy on  \emph{reliable} DRAM. This implies that introducing bit error is not a regularization technique,\footnote{Regularization is a technique that makes slight modifications to the training algorithm such that the DNN model generalizes better.} but rather, a way of obtaining congruence between the DNN training algorithms and the errors injected by approximate DRAM.

\partitle{Correcting Implausible Values}
\label{sec:aprox_fault_mitigation}
While executing curricular retraining, a single bit error in the exponent bits of a floating point value can cause \emph{accuracy collapse} in the trained DNN. For example, a bit error in the exponent of a weight creates an enormously large value (e.g., >$10^8$) that propagates through the DNN layers, dominating weights that are significantly smaller (e.g., <$10$). 

To avoid this issue, we propose a mechanism to avoid accuracy collapse caused by bit errors introduced by approximate DRAM. The key idea of our mechanism is to correct the values that are \emph{implausible}. When a value is loaded from memory, our mechanism probabilistically detects that a data type likely contains an error by comparing its value against predefined thresholds. The thresholds of the curricular retraining data types are computed during training of the baseline DNN {on} DRAM with nominal parameters. Those thresholds usually have rather small values (e.g., most weights in SqueezeNet1.1 are within the range [-5,5]). 

Upon detection of an error {(i.e, the fact that a value is out of the threshold range)} during curricular retraining, EDEN 1) corrects the erroneous value by zeroing the value, and 2) uses the corrected value for curricular retraining. 

Our mechanism for correcting implausible values can be implemented in two ways. First, a software implementation that modifies the DNN framework to include extra instructions that correct implausible values {resulting from} each DNN memory access. Second, a hardware implementation that adds a simple hardware logic to the memory controller that corrects implausible values {resulting from} each approximate DRAM memory request. Section~\ref{sec:mc} describes our low cost hardware implementation.

In our experiments, we find that our mechanism for correcting implausible values increases the tolerable bit error rate from $10^{-7}$ to $10^{-3}$ to achieve <1\% accuracy degradation in the eight FP32 DNNs we analyze. We evaluate an alternative mechanism for error correction that saturates an out-of-threshold value {(by resetting to the closest threshold value)} instead of zeroing it. We observe that saturating obtains lower {DNN} accuracy than zeroing at the same approximate DRAM bit error rate across all DNN models (e.g., 8\% on CIFAR-10 and 7\% on ImageNet). We also correct implausible values {during the execution of DNN inference to improve the} inference accuracy (Section~\ref{sec:inference}).

\subsection{DNN Error Tolerance Characterization}
\label{sec:networkcharacterization}

\eden{} aims to guarantee that the accuracy of a DNN meets the minimum value required by the user. To this end, \eden{} characterizes the boosted DNN (obtained from our boosting mechanism in Section~\ref{sec:boosting}) to find the maximum tolerable bit error rate (BER) by progressively decreasing the approximate DRAM parameters, i.e., voltage and latency.
\eden{} performs either a \textit{coarse-grained} or a \textit{fine-grained} DNN error tolerance characterization.

\partitle{Coarse-Grained Characterization}
EDEN's coarse-grained characterization determines the highest BER that can be applied uniformly to the entire DNN, while meeting the accuracy requirements of the user. This characterization is useful for mapping the DNN to commodity systems (see Section~\ref{sec:netmapping}) that apply reduced DRAM parameters to an entire DRAM {module} (without fine-grained control).

To find the highest BER that satisfies the accuracy goal, our coarse-grained characterization method performs a logarithmic-scale binary search on the error rates. We can use binary search because we found that DNN error-tolerance curves are monotonically decreasing. To adjust the BER while doing this characterization, \eden{} can either 1) tune the parameters of approximate DRAM, or 2) use DRAM error models for injecting {bit} errors {into memory locations} (see Section~\ref{sec:dramerrormodels}). \eden{} optimizes the error resiliency of a DNN by repeating cycles of DNN error tolerance boosting (Section~\ref{sec:boosting}), {coarse-grained DNN characterization, and DNN to DRAM mapping (Section~\ref{sec:netmapping})} until the highest tolerable BER stops improving. We evaluate our coarse-grained characterization mechanism in Section~\ref{sec:coarse_grain_characterization_and_mapping}. 

\partitle{Fine-Grained Characterization}
\label{sec:flatcharacterization}
EDEN can exploit variation in the error tolerances of different DNN data types by clustering the data according to its error tolerance level, and assigning each cluster to a different DRAM partition {whose error rate matches the error tolerance level of the cluster} (see Section~\ref{sec:netmapping}). For example, we find {that} the first and {the} last convolutional layers have tolerable BERs 2-3x smaller than the average middle layer in a DNN (in agreement with prior work \cite{approxnn1, approxnn2}).

To conduct a fine-grained DNN characterization, EDEN searches for the highest tolerable BER of each weight and IFM that still yields an acceptable DNN accuracy. This search space is exponential with respect to the DNN's layer count. To tackle the search space challenge, EDEN employs a DNN data sweep procedure that {performs} iterations over a list of DNN data types. The mechanism tries to increase the tolerable error rate of a data type by a {small amount}, and tests if the DNN still meets the accuracy requirements. When a DNN data type cannot tolerate more increase in error {rate}, it is removed from the sweep list. We evaluate our fine-grained characterization mechanism in Section~\ref{sec:fine_grain_characterization_and_mapping}.

\partitle{Effect of Pruning} EDEN does not include pruning (Section~\ref{sec:dnn_background}) as part of its boosting routine due to two observations. 
First, we {find} that DNN sparsification does not improve the error tolerance. Our experiments show that when we {create} 10\%, 50\%, 75\%, and 90\% sparsity through energy-aware pruning~\cite{energyawarepruning}, error tolerance of FP32 and int8 {DNNs, DNN} error tolerance does not improve significantly. Second, the zero values in the network, which {increase} with pruning, are sensitive to memory error perturbations.

\subsection{DNN {to} DRAM Mapping}
\label{sec:netmapping}

After characterizing the error tolerance of {each DNN data type}, \eden{} maps each data type to the appropriate DRAM partition (with the appropriate voltage and latency parameters) that satisfies {the} data type's error tolerance.
Our mechanism aims to map a data type that is very tolerant (intolerant) to errors into a DRAM partition with the highest (lowest) BER, matching the error tolerance of the DNN and the BER of the DRAM partition as much as possible.

\partitle{DRAM Bit Error Rate Characterization} To obtain the BER characteristics of a DRAM device (both in aggregate and for each partition), we perform reduced voltage and reduced latency tests for a number of data patterns.  
For each voltage level, we iteratively test two consecutive rows at a time. We populate these rows with inverted data patterns for the worst-case evaluation. Then, we read each bit with reduced timing parameters (e.g., tRCD). This characterization requires fine-grained control of the DRAM timing parameters and supply voltage level. \eden{}'s characterization mechanism is very similar to experimental DRAM characterization mechanisms proposed and evaluated in prior works for DRAM voltage~\cite{voltron,vampire} and DRAM latency~\cite{flydram, solardram, diva, aldram,chang2017thesis,softmc}.

\partitle{Coarse-grained {DNN to DRAM} module mapping} All DNN data types stored within the same DRAM module are exposed to the \emph{same} DRAM voltage level and timing parameters. These parameters are tuned to produce a bit error rate that is tolerable by \emph{all} DNN data types that are mapped to the module. 

Under coarse-grained mapping, the application does \emph{not} need to be modified. Algorithms used in DNN inference are oblivious to the DRAM mapping used by the memory controller. The memory controller maps all inference-related requests to the appropriate approximate DRAM module. Data that cannot tolerate bit errors at any reduced voltage and latency levels is stored in a separate DRAM module whose voltage and latency parameters follow the manufacturer specifications.

Coarse-grained mapping can be easily supported by existing systems that allow the modification of $V_{dd}$ and/or $t_{RCD/RP}$ parameters in the BIOS across the entire DRAM module. Section~\ref{sec:mc} describes the simple hardware changes required to support coarse-grained mapping. We evaluate our coarse-grained mapping mechanism in Section~\ref{sec:coarse_grain_characterization_and_mapping}.

\partitle{Fine-grained {DNN to DRAM module mapping}}
DNN data types stored in different DRAM partitions can be exposed to \emph{different} DRAM voltage levels and/or timing parameters. DRAM can be partitioned at chip, rank, bank, or subarray level granularities.  Algorithm~\ref{alg:fine-grained-mapping} describes our algorithm for fine-grained mapping of DNN data to DRAM partitions. Our algorithm uses rigorous DRAM characterization and DNN characterization to iteratively assign DNN data to DRAM partitions in three basic steps. First, our mechanism looks for DRAM partitions that have BERs lower than the tolerable BER of a given DNN data type. Second, we select the DRAM partition with the largest parameter reduction that meets the BER requirements. Third, if the partition has enough space available, our mechanism assigns the DNN data type to the DRAM partition. We evaluate our fine-grained mapping mechanism in Section~\ref{sec:fine_grain_characterization_and_mapping}.

\vspace{-0.2cm}
\begin{algorithm}[!h]
\caption{Fine-grained {DNN to DRAM} mapping}
\begin{lstlisting}[deletendkeywords={filter}]
function DNN_to_DRAM_Mapping(DNN_characterization, DRAM_characterization):
    sorted_data = sort_DNN_data(DNN_characterization)
    for (target_BER, DNN_data) in sorted_data:
        # Find the DRAM partition that has the least voltage/latency at target_BER, and can fit the DNN_data
        for DRAM_partition in DRAM_characterization
             partition_params = get_voltage_latency(DRAM_partition, target_BER)
             if DNN_data.size < DRAM_partition.size  :
                 if partition_params < best_parameters:
                     best_parameters = partition_params
                     chosen_partition = DRAM_partition
                     DRAM_partition.size -= DNN_data.size
        final_mapping[chosen_partition].append(DNN_data)
    return final_mapping
\end{lstlisting}
\label{alg:fine-grained-mapping}
 \end{algorithm}
 \vspace{-0.2cm}

A system that supports fine-grained mapping requires changes in the memory controller (for voltage and latency {adjustment}) and in DRAM (for only voltage {adjustment}). We describe the hardware changes required to support fine-grained mapping in Section~\ref{sec:mc}.

\subsection{DNN Inference with Approximate DRAM}
\label{sec:inference}

\eden{} generates a boosted DNN for running inference in a target system that uses approximate DRAM. 
\eden{} does not require any modifications in DNN inference hardware, framework, or algorithm, except for \emph{correcting implausible values}. Similar to what happens in our curricular retraining (Section~\ref{sec:boosting}), a single bit error in the exponent bits of a floating point value can cause \emph{accuracy collapse} during DNN inference. We use the same mechanism for correcting implausible values {in our} curricular retraining mechanism (i.e., {we zero the values that are outside of a predefined threshold range}) to avoid accuracy collapse caused by bit errors introduced by approximate DRAM during DNN inference.
\section{Enabling EDEN with Error Models}
\label{sec:dramerrormodels}

\eden{} requires extensive characterization of the target approximate DRAM device for boosting DNN error tolerance (Section~\ref{sec:boosting}), {characterization of} DNN error tolerance (Section~\ref{sec:networkcharacterization}), and mapping {of} the DNN to the approximate DRAM device (Section~\ref{sec:netmapping}). However, applying \eden{} in a target system {where DNN {inference} can be performed} is not always feasible or practical. For example, a low-cost DNN inference accelerator~\cite{eyeriss} might perform very slowly when executing our curricular retraining mechanism, because it is \emph{not} optimized for training. Similarly, the target hardware might not be available, or might have very limited availability (e.g., in the pre-production phase of a new approximate hardware design).

To solve this problem and enable \eden{} even when target DRAM devices are not available for characterization, we propose to execute the \eden{} framework in a  system that is different from the target approximate system. We call this idea \emph{\eden{} offloading}. The main challenge of offloading \eden{} to a different system is how to faithfully emulate the errors injected by the \emph{target} approximate DRAM into the DNN. To address this challenge, we use four different error models that are representative of most of the error patterns that are observed in real approximate DRAM modules.

\partitle{\eden{}'s {DRAM} Error Models}
\eden{} uses four probabilistic error models that closely fit the error patterns observed in a real approximate DRAM module.
Our models contain information about the location of weak cells in the DRAM module, which is used to decide the spatial distribution of bit errors during DNN error tolerance boosting. \agycomment{This sentence does not cover the fourth error model}
We create four different types of error models from the data we obtain based on our characterization of existing DRAM devices using SoftMC~\cite{softmc} and a variety of DDR3 and DDR4 DRAM modules. 
Our error models are consistent with the error patterns observed by prior works~\cite{aldram, flydram, voltron, solardram, drange, puf}.
In addition, our error models are parameterizable and can be tuned to model individual DRAM chips, {ranks}, banks, and subarrays from different vendors. 

\begin{figure}[t]
\centering
\begin{subfigure}{\columnwidth}
\centering
\includegraphics[width=\textwidth]{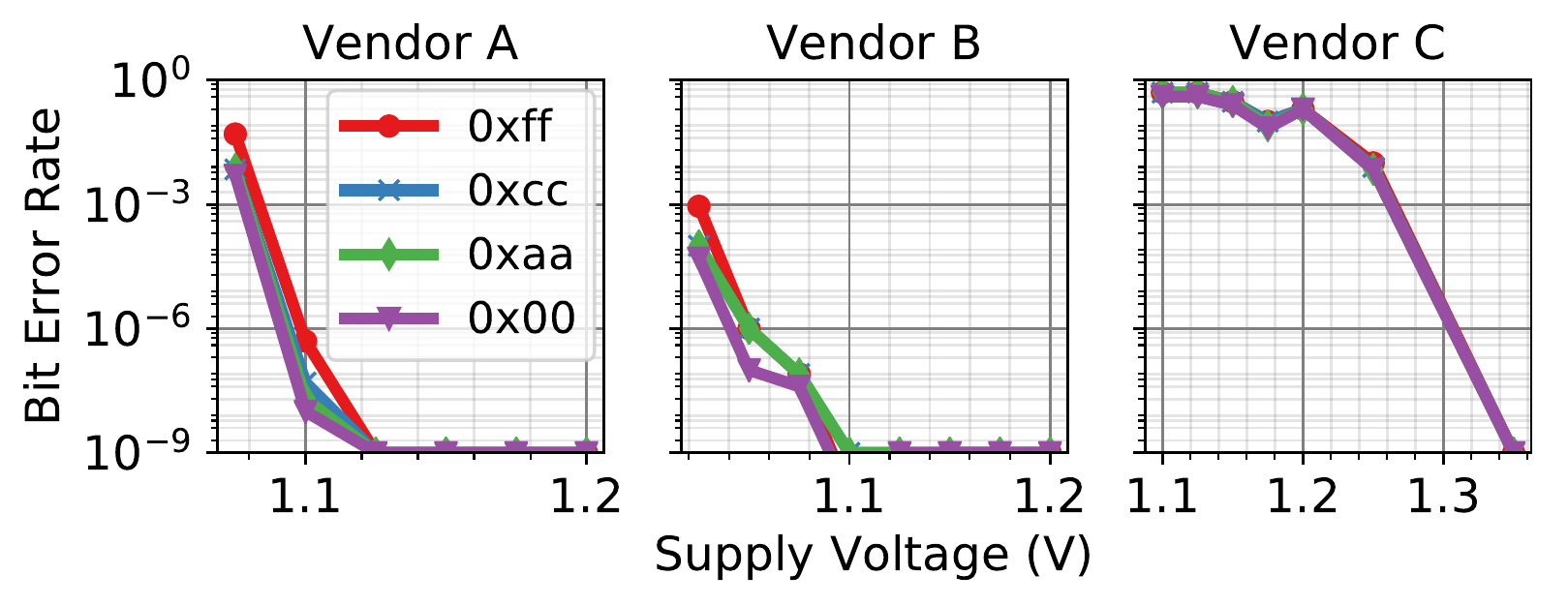}
\end{subfigure}
\begin{subfigure}{\columnwidth}
\centering
\includegraphics[width=\textwidth]{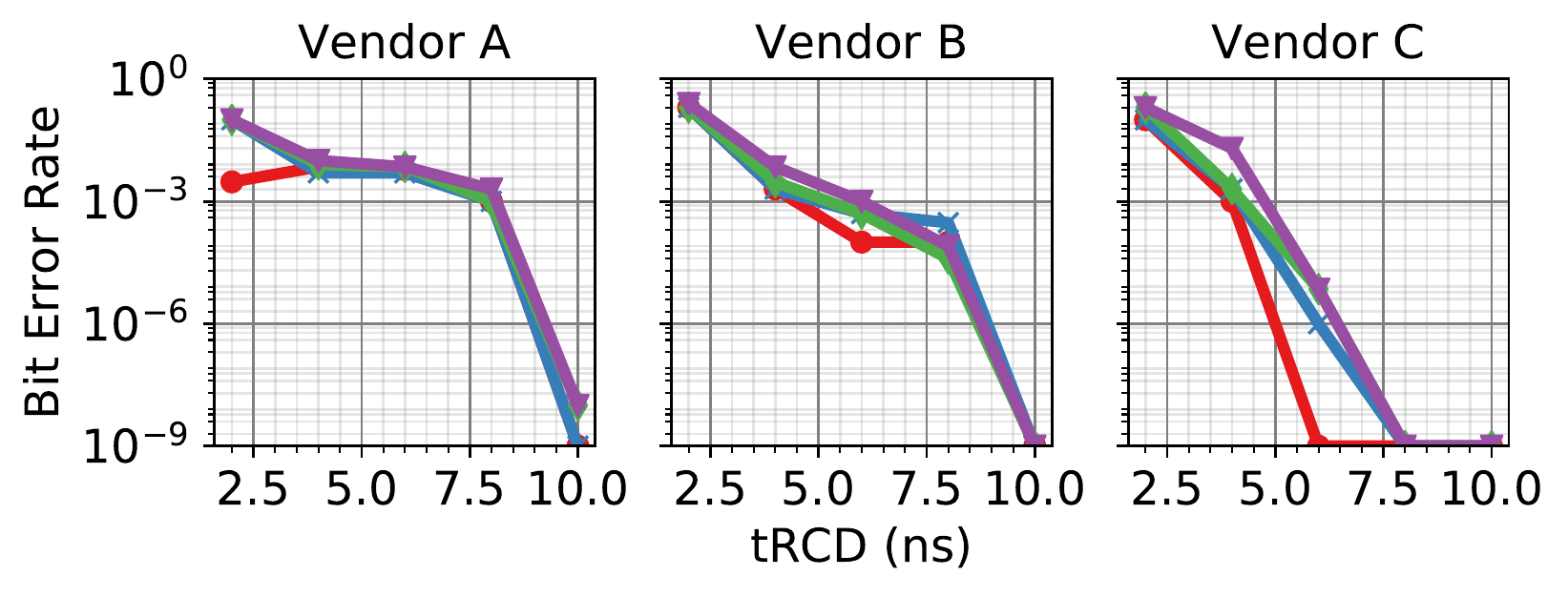}
\end{subfigure}
\vspace{-5mm}
\caption{Bit error rates depend on the data pattern stored in DRAM, with reduced supply voltage \cite{voltron} and {reduced} $t_{RCD}$~\cite{aldram, flydram, voltron, diva, solardram}, motivating Error Model 3. Data is based on DDR3 DRAM modules from three {major} vendors.}
\label{fig:vdd_curve}
\vspace{-6mm}
\end{figure}

\begin{itemize}
    \item {\bf Error Model 0: } the bit errors follow a uniform random distribution across a DRAM bank. 
    Several prior works observe that reducing activation latency ($t_{RCD}$) and precharge latency ($t_{RP}$) can cause randomly distributed bit flips due to manufacturing process variation at the level of DRAM cells~\cite{goossens, aldram, diva, flydram}. 
    We model these errors with two key parameters: 1) $P$ is the percentage of \emph{weak cells} (i.e., cells that fail with reduced DRAM parameters), and 2) $F_A$ is the probability of an error in any weak cell. Such uniform random distributions {are} already observed in prior works \cite{ontheretentiontimedistribution, avatar, rana, baek2013refresh}.

    \item {\bf Error Model 1: } the bit errors follow a vertical distribution across the bitlines of a DRAM bank.
    Prior works~\cite{solardram,diva,voltron,flydram} observe that some bitlines experience more bit flips than others under reduced DRAM parameters due to: 1) manufacturing process variation across sense amplifiers~\cite{solardram,voltron,flydram}, and 2) design-induced latency variation that arises from the varying distance between different bitlines and the row decoder~\cite{diva}.
    We model this error distribution with two key parameters: 1) $P_{B}$ is the percentage of weak cells in bitline $B$, and 2) $F_B$ is the probability of an error in the weak cells of bitline $B$. 

    \item {\bf Error Model 2: } the bit errors follow a horizontal distribution across the wordlines of a DRAM bank.
    Prior works~\cite{flydram, voltron,solardram,diva} observe that some DRAM rows experience more bit flips than others under reduced DRAM parameters due to 1) manufacturing process variation across DRAM rows~\cite{flydram, voltron,solardram}, and 
    2) design-induced latency variation that arises from the varying distance between different DRAM rows and the row buffer~\cite{diva}.
    We model this error distribution with two key parameters: 1) $P_{W}$ is the percentage of weak cells in wordline $W$, and 2) $F_W$ is the probability {of} an error in the weak cells of wordline $W$.

    \item {\bf Error Model 3: } the bit errors follow a uniform random distribution that depends on the content of the cells (i.e., this is a data-dependent error model). Figure~\ref{fig:vdd_curve} illustrates how the bit error rates depend on the data pattern stored in DRAM, for reduced voltage (top) and reduced $t_{RCD}$ (bottom). 
    We observe that 0-to-1 flips are more probable with $t_{RCD}$ scaling, and 1-to-0 flips are more probable with voltage scaling. Prior works provide rigorous analyses of data patterns in DRAM with reduced voltage~\cite{voltron} and timing parameters~\cite{flydram} that show results similar to ours.
    This error model has three key parameters: 
    1) $P$ is the percentage of weak cells, 
    2) $F_{V1}$ is the probability {of} an error in the weak cells that contain a 1 value, and 
    3) $F_{V0}$ is the probability {of} an error in the weak cells that contain a 0 value.
\end{itemize}

\partitle{Model Selection} 
EDEN applies a maximum likelihood estimation (MLE)~\cite{dsn} procedure to determine
1)  the parameters ($P$, $F_A$, $P_B$, $F_B$, $P_W$, $F_W$,  $F_{V1}$ and  $F_{V0}$) of each error model, and 
2) the error model that is most likely to produce the errors observed in the real approximate DRAM chip. 
In case two models have very similar probability of producing the observed errors, our selection mechanism chooses {Error Model 0} if possible, or one of the error models randomly otherwise. Our selection mechanism favors {Error Model 0} because we find that it the is error model that performs better. We observe that generating and injecting errors by software with {Error Model 0} in both DNN retraining and inference is 1.3x faster than injecting errors with other error models in our experimental setup. We observe that {Error Model 0} provides 1) a reasonable approximation of {Error Model 1}, if $max(F_B)-min(F_B) < 0.05$ and $P_B~\approx~P$, and 2) a reasonable approximation of {Error Model 2}, if $max(F_W)-min(F_W) < 0.05$ and $P_W~\approx~P$.

\partitle{Handling Error Variations}
Error rates and error patterns depend on two types of factors. First, factors {intrinsic} to the DRAM device. The most common intrinsic factors are caused by manufacturer \cite{voltron,refresh1}, chip, and bank variability~\cite{puf,aldram}. Intrinsic factors are established at DRAM fabrication time. Second,  factors {extrinsic} to the DRAM device that depend on environmental or operating conditions. The most common extrinsic factors are aging \cite{aging,meza2015}, data values \cite{parbor}, and temperature \cite{temperature}. Extrinsic factors can introduce significant variability in the error patterns.

\eden{} can capture intrinsic factors in the error model with a unique DRAM characterization pass. However, capturing extrinsic factors in the error model is more challenging. Our DNN models capture {three} factors extrinsic to the DRAM device. 

First, \eden{} can capture data dependent errors by generating different error models for different DNN models (i.e., different IFM and weight values in memory). For each DNN model, \eden{} stores the actual weight and IFM values in the target approximate DRAM before characterization to capture data dependencies. 

Second, \eden{} can capture temperature variations by generating different error models for the same approximate DRAM operating at different temperatures. Errors increase with higher temperatures \cite{refresh1,aldram}, so the model must match the temperature of DNN inference execution. 

Third, \eden{} can capture DRAM aging by periodically regenerating new error models. In our experiments with real DRAM modules, we find that the errors are temporally consistent and stable for days of continuous execution (with $\pm5$\degree C deviations from the profiling temperature), without requiring re-characterization. Prior works~\cite{aldram,solardram} report similar results.

We find in our evaluation that our error models are sufficiently expressive to generate a boosted DNN that executes on real approximate DRAM with minimal accuracy loss (Section~\ref{sec:softmcdemonstration}). Our four error models are also sufficiently expressive to encompass the bit-error models proposed in prior works \cite{dsn, dramerrormodel}.

\section{Memory Controller Support}
\label{sec:mc}
To obtain the most out of EDEN, we modify the memory controller to 1) correct {implausible} values {during} both curricular retraining and {DNN} inference, 2) support coarse-grained memory mapping, and 3) support fine-grained memory mapping.

\partitle{Hardware Support for Correcting {Implausible} Values}
We correct {implausible} values that cause accuracy collapse 
{during} both curricular retraining (Section~\ref{sec:boosting}) and {DNN} inference (Section~\ref{sec:inference}). Our mechanism 1) compares a loaded value to an upper-bound and a lower-bound threshold, and 2) sets the value to zero {(i.e., supplies the load with a zero result)} in case the value is out of bounds. Because {these operations are done for \emph{every}} memory access that loads a DNN value, it can cause significant performance degradation if performed in software. To mitigate this issue, we incorporate simple hardware logic in the memory controller that we call \emph{bounding logic}. Our {bounding} logic 1) compares the exponent part of the {loaded} floating point value {to DNN-specific} upper-bound and lower-bound thresholds, and 2) zeros the input value if the value is out of bounds. In our implementation, the latency of this logic is only 1 cycle and its hardware cost is negligible.

\partitle{Enabling Coarse-Grained Mapping}
Coarse-grained mapping applies the same voltage and timing parameters to the entire DRAM for executing a particular DNN workload. However, different DNN workloads might require applying different {sets} of DRAM parameters to maximize energy savings and performance. In {many} existing commodity systems, the memory controller sets the DRAM voltage and the timing parameters at start-up, and it is not possible to change them at runtime. To overcome this limitation, the memory controller requires minimal hardware support for changing the DRAM parameters of {each} DRAM {module} at runtime. 

\partitle{Enabling Fine-Grained Mapping}
Fine-grained mapping applies different voltage and/or timing parameters to different DRAM partitions. 

To apply different voltages to different memory partitions, \eden{} 1) adopts the approach used by Voltron \cite{voltron} to implement a robust design for voltage scaling at the bank granularity based on modest changes to the power delivery network, and 2) tracks which memory partition is operating at what voltage. To implement this mechanism in commodity DDR4/LPDDR4 chips with 16/32 banks, \eden{} requires at most 32B of meta-data to represent all 8-bit voltage {step} values.

To apply different timing parameters to different memory partitions, \eden{} requires memory controller support for 1) {configuring} the target memory partition to {operate at} specific timing parameters, and 2) tracking which memory partition is operating at what latency. For the timing parameter we tested in our evaluation ($t_{RCD}$), 4-bits are enough to encode {all possible values} of the parameter with enough resolution.

It {is sufficient} for \eden{} to split DRAM into at most $2^{10}$ partitions, because most commonly used DNN architectures have at most 1024 different types of error-resilient IFMs and weights. \eden{} requires 1KB of metadata to support $2^{10}$ partitions. To support mappings at subarray level granularity (i.e., the finest supported granularity), \eden{} needs a larger amount of metadata. For example, for an 8GB DDR4 DRAM module {with 2048 subarrays}, \eden{} needs to store 2KB of metadata.

\section{DNN Accuracy Evaluation}
\label{sec:accuracy_eval}

In this section, we evaluate EDEN's ability to improve DNN accuracy in approximate DRAM. We explain our methodology (Section~\ref{sec:accuracy_methodology}), evaluate the accuracy of our error models (Section~\ref{sec:error_model_validation}), evaluate the error tolerance of the DNN baselines (Section~\ref{sec:baselineerrorresilience}), and analyze the accuracy of our curricular retraining mechanism (Section~\ref{sec:softmcdemonstration}).

\subsection{Methodology}
\label{sec:accuracy_methodology}
We use an FPGA-based infrastructure running SoftMC~\cite{softmc,SoftMCGithub} to reduce DRAM voltage and timing parameters. SoftMC allows executing memory controller commands on individual banks, and modifying $t_{RCD}$ and other DRAM timing parameters. We perform all our experiments at room temperature. Using this infrastructure, we can obtain characteristics of real approximate DRAM devices. However, our infrastructure also has some performance limitations caused by delays introduced with SoftMC's FPGA buffering, host-FPGA data transmission, and instruction batching on the FPGA. 

To overcome these performance limitations, we emulate real approximate DRAM modules by using the error models described in Section~\ref{sec:dramerrormodels}. To ensure that our evaluation is accurate, we validate our error models against real approximate DRAM devices (Section~\ref{sec:error_model_validation}).

\begin{figure}[h]
\centering
\vspace{-1.3mm}
\includegraphics[width=0.45\textwidth]{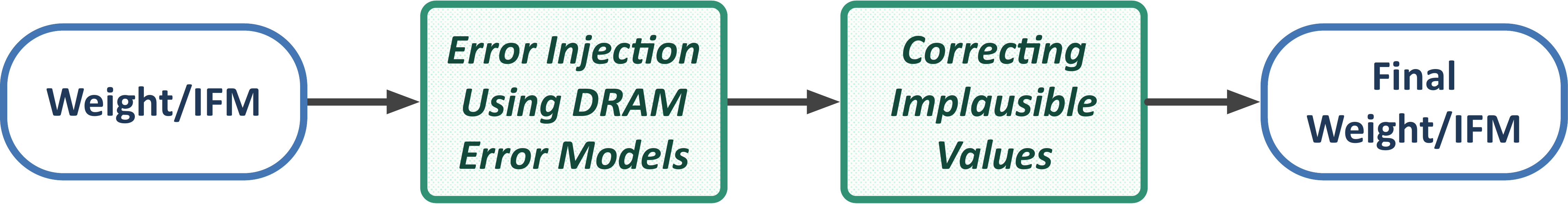}
\vspace{-3mm}
\caption{Methodology to incorporate DRAM error models in the DNN evaluation framework.} 
\label{fig:errorflow}
\vspace{-5.2mm}
\end{figure}

We incorporate EDEN's error models into DNN inference libraries by following the methodology described in Figure~\ref{fig:errorflow}. We create a framework on top of PyTorch \cite{pytorch} that allows us to modify the loading of weights and IFMs. Our {PyTorch} implementation 1) injects errors {into} the original IFM and weight values {using} our DRAM error models, and 
2) applies our mechanism to correct implausible values caused by bit errors in IFMs and weights (Section~\ref{sec:boosting}).
Our DRAM error models are implemented as custom GPU kernels for efficient and simple integration into PyTorch. This simulation allows us to obtain DNN accuracy estimates 80-90x faster than with the SoftMC infrastructure.

\partitle{DNN Baselines}
\label{sec:baselines}
We describe the DNN baselines that we use in the evaluation of the three \eden{} steps (Sections~\ref{sec:boosting},~\ref{sec:networkcharacterization}, and \ref{sec:netmapping}).
Table~\ref{table:networks} lists the eight modern and commonly-used DNN models we evaluate.  We target both small (e.g., CIFAR-10~\cite{cifar10}) and large-scale (e.g., ILSVRC2012~\cite{ilsvrc})  image classification datasets. ResNet101~\cite{resnet}, VGG-16~\cite{vgg}, and DenseNet201~\cite{densenet} models are top-five winners of past ImageNet ILSVRC competitions \cite{ilsvrc, imagenet2}. We use Google MobileNetV2~\cite{mobilenetv2} to test smaller, mobile-optimized networks that are widely used on mobile platforms, and \sloppy SqueezeNet~\cite{squeezenet} to test embedded, real-time applications. Table~\ref{table:networks} also shows the summed sizes of all IFMs and weights \roki{of} each network for processing one input, which is a good indicator of the memory intensity of each DNN model.

\begin{table}[h]
\centering
\vspace{-0.3cm}
\small
\begin{tabular}{ll|rr}

\bf{Model}                      & \bf{Dataset}             & \bf{Model Size} & \bf{\thead{\roki{IFM+Weight}\\Size}} \\ \hline
ResNet101~\cite{resnet}         & CIFAR10~\cite{cifar10}   & 163.0MB         & 100.0MB         \\
MobileNetV2~\cite{mobilenetv2}  & CIFAR10~\cite{cifar10}   & 22.7MB          & 68.5MB        \\
VGG-16~\cite{vgg}               & ILSVRC2012~\cite{ilsvrc} & 528.0MB         & 218.0MB         \\
DenseNet201~\cite{densenet}     & ILSVRC2012~\cite{ilsvrc} & 76.0MB          & 439.0MB         \\
SqueezeNet1.1~\cite{squeezenet} & ILSVRC2012~\cite{ilsvrc} & 4.8MB           & 53.8MB        \\ 
Alexnet~\cite{imagenet2}        & CIFAR10~\cite{cifar10}   & 233.0MB         & 208.0MB         \\ \hline
YOLO~\cite{yolo}                & MSCOCO~\cite{coco}       & 237.0MB         & 360.0MB         \\
YOLO-Tiny~\cite{yolo}           & MSCOCO~\cite{coco}       & 33.8MB          & 51.3MB        \\ \hline
{LeNet${}^\star$~\cite{lecun1998gradient}} & CIFAR10~\cite{cifar10} & 1.65MB & 2.30MB \\
\end{tabular}
\begin{tablenotes}
      \small
      \centering
      \item  ${}^\star$ we use this small model in {some evaluations where} the experimental setup does not support large models.
\end{tablenotes}
\vspace{0.5em}
\caption{DNN models used in our evaluations. The listed total model \roki{size} and summed IFM+weight sizes are for the FP32 variant {of each model}.}
\vspace{-0.8cm}
\label{table:networks}
\end{table}

Table~\ref{table:baselines} shows the accuracy we obtain in our experiments for our baseline networks across four different numeric precisions (int4, int8, int16 and FP32), using \emph{reliable} commodity DRAM.  We quantize using the popular symmetric linear DNN quantization scheme \cite{quantization_scheme}. This quantization scheme applies weight-dependent affine scaling to linearly map weights into {the} range $[-2^{b-1},\ 2^{b-1}-1]$, where $b$ is the target model weight bit precision. YOLO and YOLO-Tiny's framework only support int8 and FP32 numeric precisions.

\begin{table}[h]
\small
\centering
\begin{threeparttable}
\begin{tabular}{l|rrrr}
\bf{Model}          & \bf{int4}      & \bf{int8} & \bf{int16} & \bf{FP32} \\ 
\hline
ResNet101~\cite{resnet}         & 89.11\% & 93.14\% & 93.11\% & 94.20\% \\ 
MobileNetV2~\cite{mobilenetv2}  & 51.00\% & 70.44\% & 70.46\% & 78.35\% \\ 
VGG-16~\cite{vgg}               & 59.05\% & 70.48\% & 70.53\% & 71.59\% \\ 
DenseNet201~\cite{densenet}     & 0.31\%  & 74.60\% & 74.82\% & 76.90\% \\ 
SqueezeNet1.1~\cite{squeezenet} & 8.07\%  & 57.07\% & 57.39\% & 58.18\% \\ 
Alexnet~\cite{imagenet2}        & 83.13\% & 86.04\% & 87.21\% & 89.13\% \\ \hline 
YOLO${}^\star$~\cite{yolo}      & -- & 44.60\% & -- & 55.30\% \\ 
YOLO-Tiny${}^\star$~\cite{yolo} & -- & 14.10\% & -- & 23.70\% \\ \hline
{LeNet~\cite{lecun1998gradient}} & -- & 61.30\% & -- & 67.40\% \\
\end{tabular}
\centering
\begin{tablenotes}
      \small
      \centering
      \item  ${}^\star$ these models use mean average precision (mAP)  instead of the accuracy metric.
\end{tablenotes}
\end{threeparttable}
\caption{Baseline accuracies of the networks used in our evaluation with reliable DRAM memory (no bit errors) using different numeric precisions.}
\vspace{-0.9cm}
\label{table:baselines}
\end{table}

Our baseline accuracies match stated numbers in relevant literature \cite{resnet, mobilenetv2, vgg, densenet, squeezenet}. Two of the models, DenseNet201 and SqueezeNet1.1, suffer from accuracy collapse at 4-bit precision. We did not use hyper-parameter tuning in our baselines or subsequent experiments. All results use the default DNN architectures and learning rates.

\subsection{Accuracy Validation of the Error Models}
\label{sec:error_model_validation}

EDEN uses errors obtained from real DRAM devices to build and select accurate error models. 
We profile the DRAM 1) before running DNN inference, and 2) when the environmental factors that can affect the error patterns change (e.g., when temperature changes). We find that an error model can be accurate for many days if the environmental conditions do not change significantly, as also observed in prior work~\roki{\cite{solardram, aldram, diva}}. 

We derive our probabilistic error models (Section~\ref{sec:dramerrormodels}) from data obtained from eight real DRAM modules. We use the same FPGA infrastructure as the one described in Section~\ref{sec:accuracy_methodology}.
We find that complete profiling of a  16-bank, 4GB DDR4 DRAM module takes under 4 minutes in our evaluation setup. We can speed up the profiling time by 2-5x using more sophisticated DRAM profiling methodologies~\cite{reaper}.

\begin{figure}[h]
 \centering
 \vspace{-0.4cm}
 \includegraphics[width=0.42\textwidth]{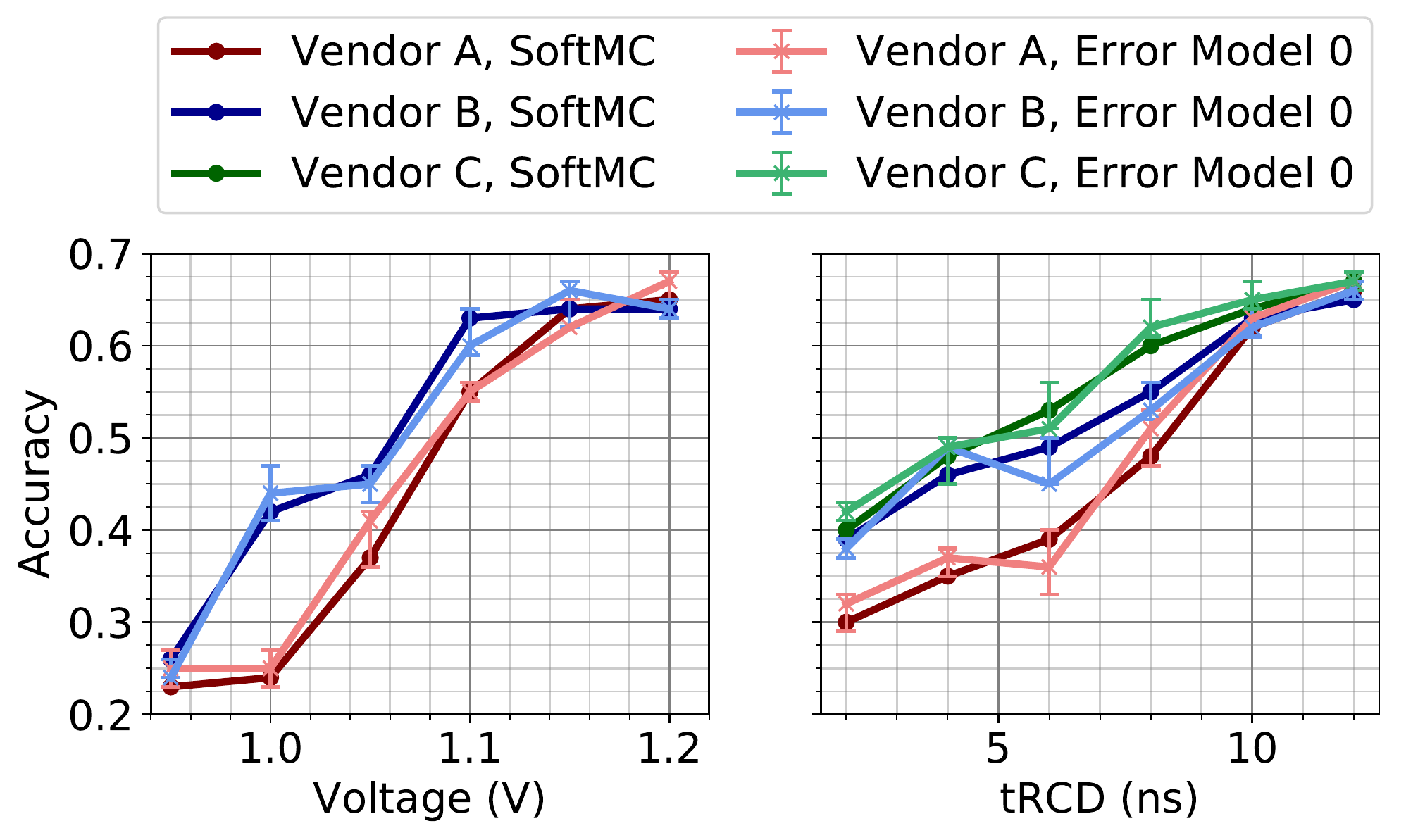}
 \vspace{-0.4cm}
  \caption{LeNet/CIFAR-10 accuracies obtained using real approximate DRAM devices ({via} SoftMC) and using our {Error Model 0}. Error bars show the 95\% confidence interval of {Error Model 0}.}
 \label{fig:softmc_error_model}
 \vspace{-0.5cm}
 \end{figure}

We validate our error models by comparing the DNN accuracy obtained after injecting bit errors using our {DRAM} error models to the accuracy obtained with each real approximate DRAM module. Figure~\ref{fig:softmc_error_model} shows an example of the DNN accuracy obtained using DRAM modules from {three major} vendors with reduced voltage and $t_{RCD}$, and the DNN accuracy obtained using our Error Model 0. We use Error Model 0 because it is the model that fits better the errors observed in the three tested DRAM modules.
Our main observation is that the DNN accuracy obtained with our model is very similar to that {obtained} with real approximate DRAM {devices}. We conclude that our error models mimic very well the errors observed in real approximate DRAM {devices}.

\subsection{Error Tolerance of Baseline DNNs}
\label{sec:baselineerrorresilience}
To better understand the baseline error tolerance of each DNN (before boosting the error tolerance), we examine the error tolerance of the baseline DNNs. This also shows us how differences in quantization, best-fit error model, and BER can \roki{potentially} affect the final DNN accuracy. 

\begin{figure}[h]
\centering
\vspace{-0.3cm}
\includegraphics[width=0.8\linewidth]{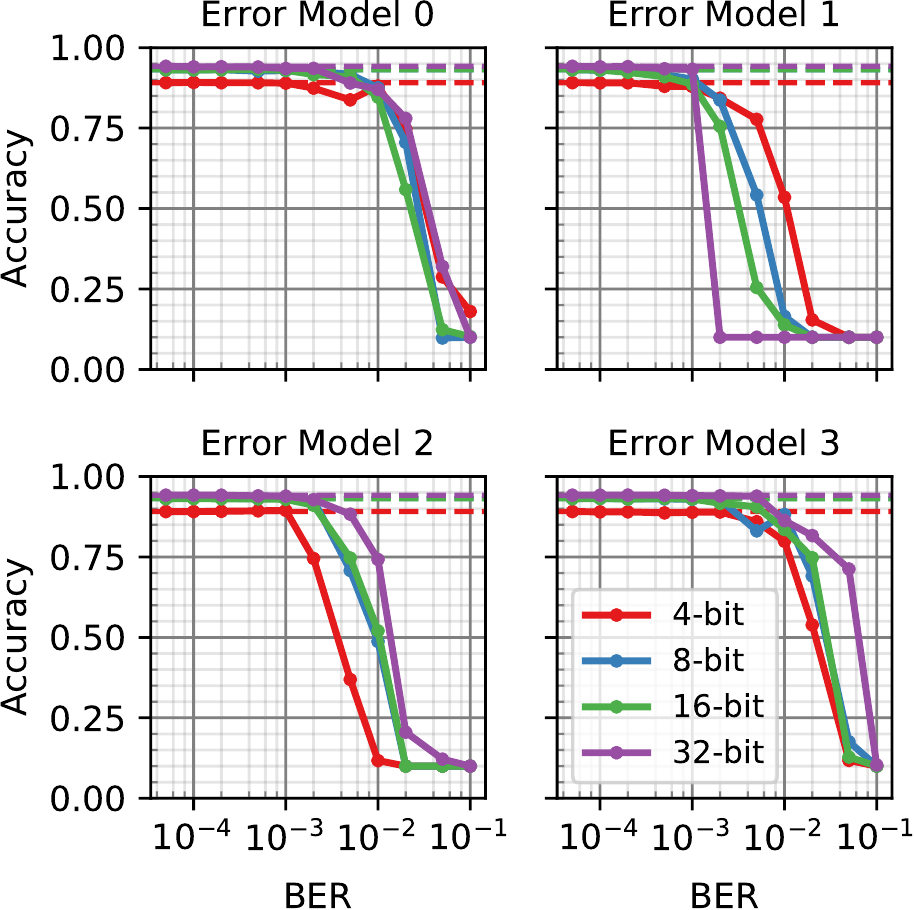}
\vspace{-0.4cm}
\caption{ResNet101 accuracy across different BERs ($x$-axis) and quantization levels when we use four error models to inject bit errors. We fit the parameters of the error models to the errors {observed by} reducing $tRCD$ in a real DRAM device from {Vendor A.}}
\label{fig:error_models}
\vspace{-0.5cm}
\end{figure}

Figure~\ref{fig:error_models} shows the accuracy of ResNet101 at different \roki{precision levels} and BERs using all four error models. We see that all DNNs exhibit an accuracy drop at high BER ($>10^{-2}$), but different error models cause the drop-off for all DNNs to be higher or lower. This is rooted in how each error model disperses bit errors into the DNN IFMs and weights. A good example of this is Error Model 1, which exhibits the most early and extreme drop-offs, especially for FP32 DNNs. 
We find that the cause of this is that, in our experimental setup, IFMs and weights are aligned in DRAM, so the MSBs of different DNN data types are mapped to the same bitline B. If the percentage of weak cells in bitline B ($P_B$) is high, the DNN suffers many MSB failures.
However, Error Model 0 distributes these weak cell failures uniformly and randomly across the bank, causing far fewer MSB failures. In general, the way in which each error model captures the distribution of weak cells across data layout \roki{in memory} greatly affects its impact on the error curve. 

\partitle{Quantization} Precision also affects the error model and the error tolerance curve. For example, in Error Model 2, we observe that the int-4 DNN has the weakest error tolerance curve. We find that this is because Error Model 2 clusters weak cells along a row: a large number of neighboring 4-bit values end up corrupted when Error Model 2 indicates a weak wordline. This is in contrast to larger precisions, which might have numbers distributed more evenly across rows, or error models that do not capture error locality (e.g., Error Model 0). In general, we find that clusters of erroneous values cause significant problems with accuracy (the errors compound faster as they interact with each other in the DNN). Such locality of errors is more common in low-bitwidth precisions and with spatial correlation-based error models (Error Models 1 and 2).

\partitle{DNN Size} We observe that larger DNNs (e.g., VGG16) are more error resilient. Larger models exhibit an accuracy drop-off at higher BER ($>10^{-2}$) as compared to smaller models (e.g. SqueezeNet1.1, $<10^{-3}$). These results are not plotted.

\partitle{Accuracy Collapse} We can observe the accuracy collapse phenomenon caused by implausible values (see Section~\ref{sec:boosting}) when we increase the bit error rate over $10^{-6}$ in large networks. These implausible values propagate, and in the end, they cause accuracy collapse in the DNN.

\subsection{Curricular Retraining Evaluation}
\label{sec:softmcdemonstration}

We run DNN inference on real DRAM devices using the boosted DNN model generated by our curricular retraining mechanism. {To our knowledge, this is the first demonstration of DNN inference on real approximate memory. We also evaluate our curricular retraining mechanism using our error models (see Section~\ref{sec:dramerrormodels}).}

\partitle{Experimental Setup}
{We evaluate curricular retraining using real DRAM devices by running LeNet~\cite{lecun1998gradient} on the CIFAR-10~\cite{cifar10} validation dataset. We use SoftMC~\cite{softmc} to scale {$V_{DD}$} and $t_{RCD}$ on an FPGA-based infrastructure connected to a DDR3 DRAM module. {We also evaluate curricular retraining using our error models by running ResNet~\cite{resnet} on the CIFAR-10 validation dataset.}}

\begin{figure}[h]
\centering
\vspace{-0.3cm}
\includegraphics[width=0.45\textwidth]{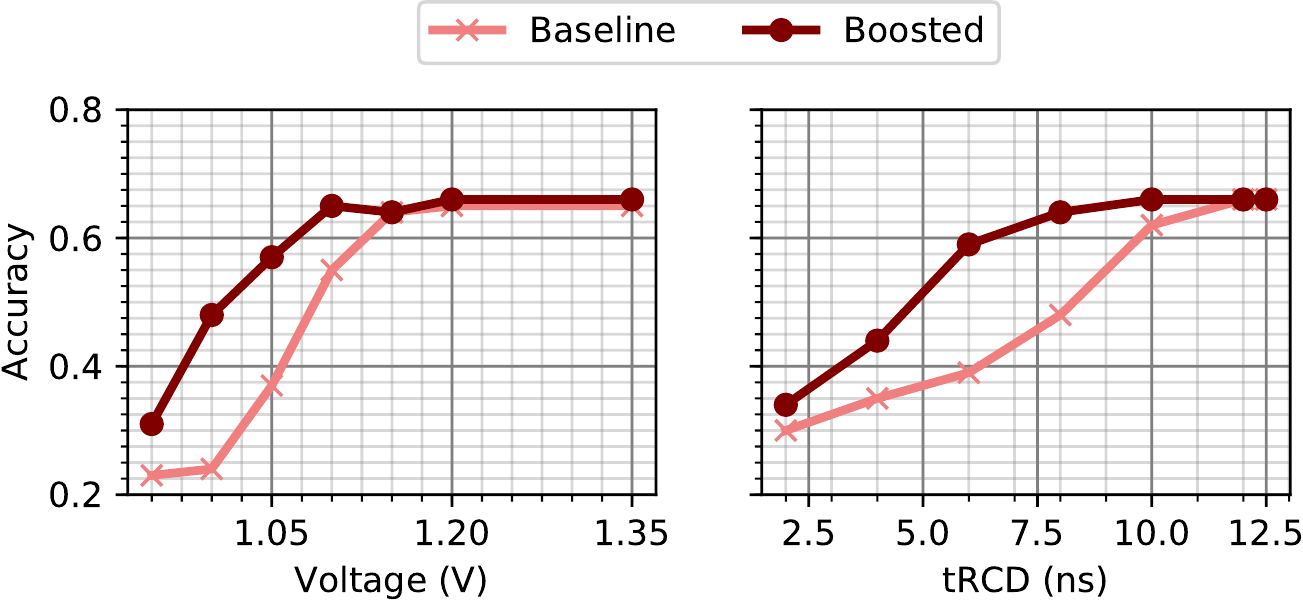}
\vspace{-0.4cm}
\caption{LeNet accuracy using baseline and boosted DNNs.}
\vspace{-0.4cm}
\label{fig:softmc_dnn}
\end{figure}

\partitle{{Results with Real DRAM}}
Figure~\ref{fig:softmc_dnn} shows the accuracy of 1) \lois{baseline} LeNet  without applying any retraining mechanism (Baseline), and 2) LeNeT boosted with our curricular retraining mechanism (Boosted), as a function of DRAM supply voltage and $t_{RCD}$. We make two observations. First, EDEN's boosted LeNet allows a voltage reduction of \lois{$\sim$0.25V} and a $t_{RCD}$ reduction of \lois{4.5ns}, while maintaining accuracy values equivalent to those provided by nominal voltage \lois{(1.35V)} and nominal $t_{RCD}$ \lois{(12.5ns)}. Second, the accuracy of \lois{baseline} LeNet decreases very quickly when reducing voltage and $t_{RCD}$ below the nominal values. We conclude that our curricular retraining mechanism can effectively boost the accuracy of LeNeT \lois{on} approximate DRAM with reduced voltage and $t_{RCD}$.

\begin{figure}[h]
\centering
\includegraphics[width=0.45\textwidth]{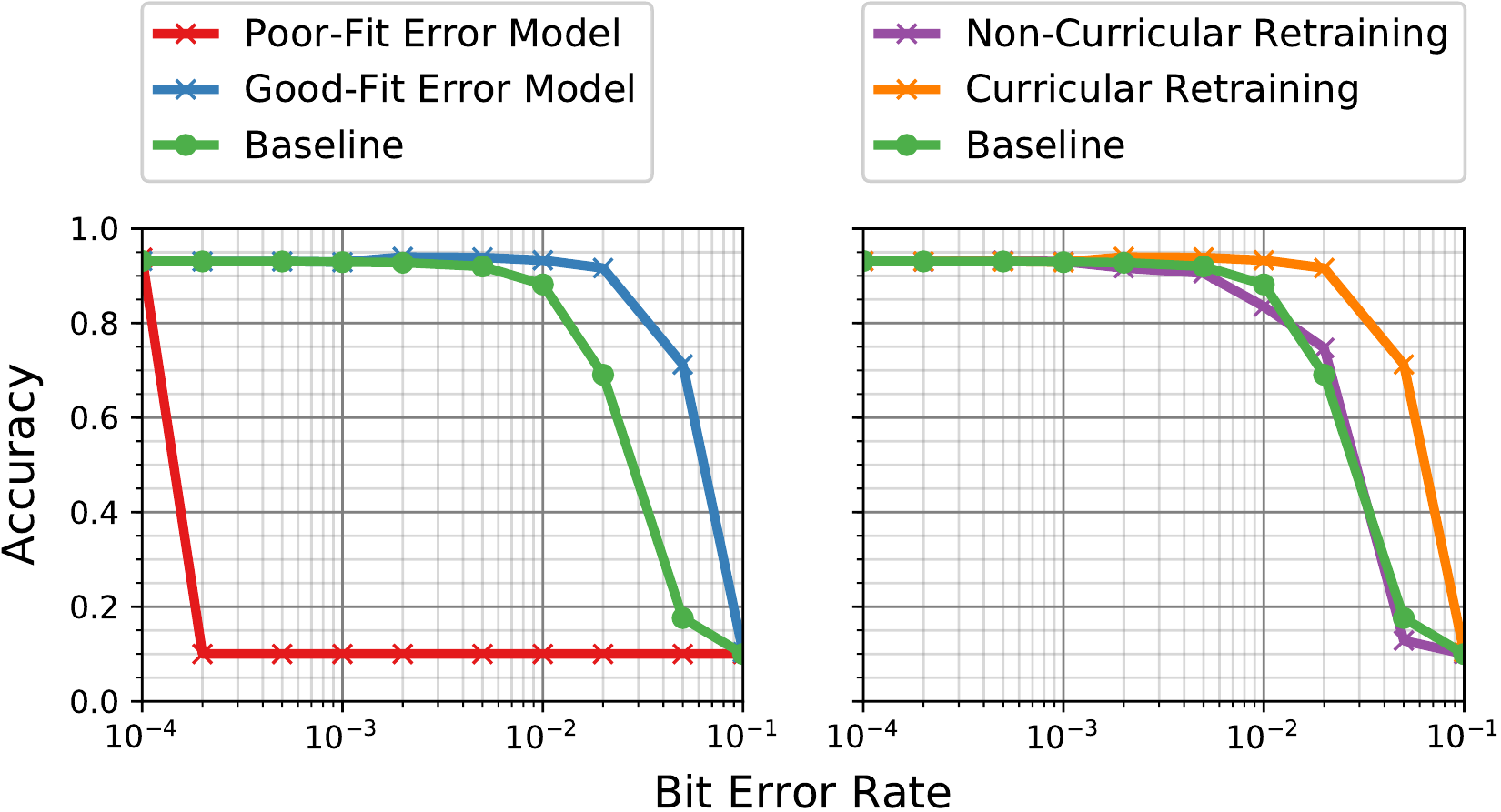}
\vspace{-0.2cm}
\caption{Accuracy of boosted ResNet101 DNNs in presence of memory errors. Left: accuracy of poor-fit and good-fit error models. Right: accuracy of non-curricular and curricular retraining using a good-fit error model.}
\label{fig:retraining}
\vspace{-0.47cm}
\end{figure}

\partitle{{Results with Error Models}}
Figure~\ref{fig:retraining} (left) shows an experiment that retrains ResNet101 with two different models: 1) a good-fit error model (that closely matches the tested device) and 2) a poor-fit error model. We make two observations. First, retraining using a poor-fit error model (red), yields little improvement over the baseline (no retraining, green). Second, retraining with a good-fit error model (blue) improves BER at the 89\% accuracy point by >10x (shifting the BER curve right). {We conclude that using a good-fit error model in the retraining mechanism is critical to avoid accuracy collapse.}

Figure~\ref{fig:retraining} (right) shows the effectiveness of our curricular retraining mechanism using a good-fit error model. We make two observations. First, the accuracy of the DNN with regular retraining (purple) collapses, compared to \lois{the} baseline DNN (no retraining, green). Second, the DNN trained with our curricular retraining (orange) exhibits a boosted error tolerance. We conclude that our curricular retraining mechanism is effective at boosting the DNN accuracy in systems that use approximate DRAM.

Running this retraining process for 10-15 epochs is sufficient to boost tolerable BERs by 5-10x to achieve the same DNN accuracy as the baseline DNN executed in DRAM with nominal parameters\onurcomment{as what?}. For our ResNet101 on CIFAR-10 with \roki{an} NVIDIA Tesla P100, this one-time boosting completes within 10 minutes.

\subsection{Coarse-Grained DNN Characterization \\and Mapping}
\label{sec:coarse_grain_characterization_and_mapping}

In this section, we show the results of \eden{}'s coarse-grained DNN characterization {(see Section~\ref{sec:flatcharacterization})} and how the target DNN model maps to an approximate DRAM with optimized parameters for a target accuracy degradation \roki{of $<1\%$}.

\partitle{Characterization}  Table~\ref{table:blanketresults} shows the DNN's maximum tolerable BER for eight DNN models with FP32 and int8 numeric precisions.

\begin{table}[h]
\vspace{-0.3cm}
\small
\centering
\setlength\tabcolsep{3pt} 
\begin{tabular}{l|ccc|ccc}
\multicolumn{1}{c}{} & \multicolumn{3}{c}{{\bf FP32}} & \multicolumn{3}{c}{{\bf int8}} \\
\textbf{Model}&BER&\textbf{$\Delta V_{DD}$}&\textbf{$\Delta t_{RCD}$} &BER&\textbf{$\Delta V_{DD}$}&\textbf{$\Delta t_{RCD}$} \\ 
\hline
\textbf{ResNet101}        & 4.0\%        & -0.30V & -5.5ns & 4.0\% & -0.30V & -5.5ns\\ 
\textbf{MobileNetV2}      & 1.0\%          & -0.25V  & -1.0ns & 0.5\% & -0.10V & -1.0ns\\ 
\textbf{VGG-16}          & 5.0\%        & -0.35V & -6.0ns  & 5.0\% & -0.35V & -6.0ns \\ 
\textbf{DenseNet201}      & 1.5\%        & -0.25V  & -2.0ns & 1.5\% & -0.25V & -2.0ns\\ 
\textbf{SqueezeNet1.1}    & 0.5\%        & -0.10V  & -1.0ns & 0.5\% & -0.10V & -1.0ns\\ 
\textbf{AlexNet}          & 3.0\%        & -0.30V  & -4.5ns & 3.0\% & -0.30V & -4.5ns\\ 
\textbf{YOLO}           & 5.0\%        & -0.35V  & -6.0ns & 4.0\% & -0.30V & -5.5ns \\ 
\textbf{YOLO-Tiny}      & 3.5\%        & -0.30V  & -5.0ns & 3.0\% & -0.30V & -4.5ns\\ 
\end{tabular}
\vspace{0.1cm}
\caption{Maximum tolerable BER for each DNN using \eden{}'s coarse-grained characterization, and DRAM parameter reduction to achieve the maximum tolerable BER. Nominal parameters are $V_{DD}=1.35V$ and $t_{RCD}=12.5ns$.}
\vspace{-0.7cm}
\label{table:blanketresults}
\end{table}

We observe that the {maximum} tolerable BER {demonstrates significant variation} depending on the DNN model. For example, YOLO tolerates 5\% BER and {SqueezeNet} tolerates only 0.5\%. We conclude that 1) the maximum tolerable BER highly depends on the DNN model, and 2) DNN characterization is required to optimize approximate DRAM parameters for each DNN model. 

\partitle{Mapping} \eden{} maps {each} DNN model to an approximate DRAM {module} that operates with the maximum reduction in voltage ($\Delta V_{DD}$) and $t_{RCD}$ ($\Delta t_{RCD}$) that leads to a BER below the maximum DNN tolerable BER {for that DNN model}. Table~\ref{table:blanketresults} shows the maximum reduction in DRAM voltage ($\Delta V_{DD}$) and $t_{RCD}$ ($\Delta t_{RCD}$) that causes a DRAM BER below the maximum tolerable BER, for a target DRAM module from vendor A. The nominal DRAM parameters for this DRAM module are $V_{DD}=1.35V$ and $t_{RCD}=12.5ns$ . We make two observations. 
First, the tolerable BER of a network is directly related to {the maximum tolerable $V_{DD}$ and $t_{RCD}$ reductions.}
Second, the reductions in $V_{DD}$ and $t_{RCD}$ are very significant compared to the nominal values. For example, \eden{} can reduce {voltage by} 26\%  and {$t_{RCD}$ by} 48\% in YOLO while maintaining the {DNN} accuracy {to be} within 1\% of the original {accuracy}.

\subsection{Fine-Grained DNN Characterization \\and Mapping}
\label{sec:fine_grain_characterization_and_mapping}
\partitle{Characterization} We characterize the ResNet101 DNN model with our fine-grained DNN characterization procedure (see Section~\ref{sec:networkcharacterization}). 
For each IFM and weight, we iteratively increase the bit error rate until we reach the maximum tolerable BER of the data type for a particular target accuracy degradation. We {perform} a full network retraining {in} each iteration.
To reduce the runtime of our procedure , we sample 10\% of the validation set during each inference run to obtain the accuracy estimate. We also bootstrap the BERs to the BER found in coarse-grained DNN characterization and use a linear scale in 0.5 increments around that value. For ResNet101, this one-time characterization completes in one hour using an Intel Xeon CPU E3-1225~\cite{xeon}. 

\begin{figure}[h]
\centering
\vspace{-0.3 cm}
\includegraphics[width=0.47\textwidth]{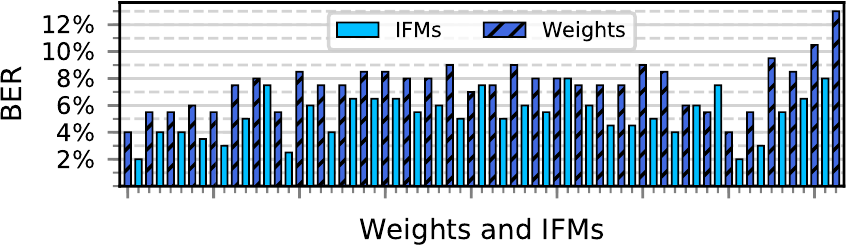}
\vspace{-0.4cm}
\caption{Fine-grained characterization of the tolerable BERs of ResNet101 IFMs and weights. \lois{Deeper} layers \lois{are} on the right.}
\label{fig:fine_grained_characterization}
\vspace{-0.6cm}
\end{figure}

Figure~\ref{fig:fine_grained_characterization} shows the maximum tolerable BER for each IFM and weight in ResNet101 obtained with our fine-grained DNN characterization method (Section~\ref{sec:networkcharacterization}), assuming a maximum accuracy loss of $<$1\%. Each bar in the figure represents the BER tolerance of an IFM or weight, and they are ordered by their {depth} in the DNN, going deeper from left to right. We make three observations. First, fine-grained characterization enables individual IFMs and weights to tolerate up to 3x BER (13\% for the last weight) of the maximum tolerable BER of the coarse-grained approach (4\% for ResNet101 in Table~\ref{table:blanketresults}). Second, weights usually tolerate more errors than IFMs. Third, the maximum tolerable BER is smaller in the first layers than in the middle layers of the DNN.
We conclude that fine-grained DNN characterization enables a significant increase in the maximum tolerable BER compared to coarse-grained characterization.

\begin{figure}[h]
\centering
\includegraphics[width=0.47\textwidth]{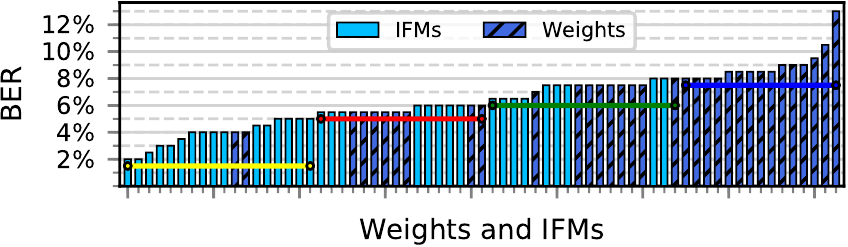}
\vspace{-0.4cm}
\caption{Mapping of ResNet101 IFMs and weights into four partitions with different $V_{DD}$ {values} ({colored horizontal lines})}
\label{fig:fine_grained_mapping}
\vspace{-0.4cm}
\end{figure}

\partitle{Mapping} We map each individual IFM or weight into different DRAM partitions based on 1) the BER {tolerance} of {each IFM and weight}, and 2) the BER of each DRAM partition, {using} our algorithm in Section~\ref{sec:netmapping}. Figure~\ref{fig:fine_grained_mapping} shows an example that maps the ResNet101 IFMs and weights from Figure~\ref{fig:fine_grained_characterization} into 4 different DRAM partitions with different voltage parameters that introduce different BERs (four horizontal colored bars), following the algorithm in Section~\ref{sec:netmapping}. 

We {conclude that the wide range of tolerable BERs across all ResNet101 data types enables the use of both 1) DRAM partitions with significant voltage reduction (e.g., horizontal red line), and 2) DRAM partitions with moderate voltage reduction (e.g., horizontal blue line). }

\section{System Level Evaluation}
\label{sec:evaluation}

We evaluate EDEN in three different DNN inference architectures: CPUs, GPUs, and inference accelerators.

\subsection{CPU Inference}
\label{sec:cpuinference}
\partitle{Experimental Setup}
We evaluate EDEN on top of a multi-core OoO CPU using the simulated core configuration listed in Table 4.  We use ZSim \cite{zsim} and Ramulator \cite{ramulator} to simulate the core and the DRAM subsystem, respectively. We use DRAMPower~\cite{drampower} to estimate \energy{} consumption for DDR4 devices. We use a 2-channel, 32-bank 8GB DDR4-2133 DRAM device. 

 \begin{table}[h]
 \small
 \begin{center}
  \vspace{-0.25cm}
 \begin{tabular}{r|p{6cm}}
 \textbf{Cores} & 2 Cores @ 4.0 GHz, 32nm, 4-wide OoO,\\
                 & Buffers: 18-entry fetch, 128-entry decode,\\
                 & 128-entry reorder buffer,  \\ 
 \textbf{L1 Caches} & 32KB, 8-way, 2-cycle, Split Data/Instr. \\
 \textbf{L2 Caches} & 512KB per core, 8-way, 4-cycle, Shared Data/Instr.,\\
                     & Stream Prefetcher \\ 
 \textbf{L3 Caches} & 8MB per core, 16-way, 6-cycle, Shared Data/Instr.,\\
                     & Stream Prefetcher \\ 
  \textbf{Main Memory} & 8GB DDR4-2133 DRAM, 2 channels, 16 banks/channel \\ 
 \end{tabular}
 \end{center}
 \caption{Simulated system configuration.}
 \label{table:zsimconfig}
 \vspace{-0.8cm}
 \end{table}
 
 We use twelve different inference benchmarks: eight from the Intel OpenVINO toolkit \cite{openvino} and four from the AlexeyAB-fork of the DarkNet framework \cite{darknet}. For each DNN, we study the FP32 and the \texttt{int8}-quantized variant. We use 8-bit quantization in our baselines, because it is commonly used for production CPU workloads \cite{quantization2}. 
 We evaluate EDEN's coarse-grained {DNN} characterization procedure and target a $<1\%$ accuracy degradation. Table~\ref{table:blanketresults} lists the reduced $V_{DD}$ and $t_{RCD}$ values. 
 
\begin{figure}[h]
\centering
\vspace{-0.4cm}
\includegraphics[width=0.48\textwidth]{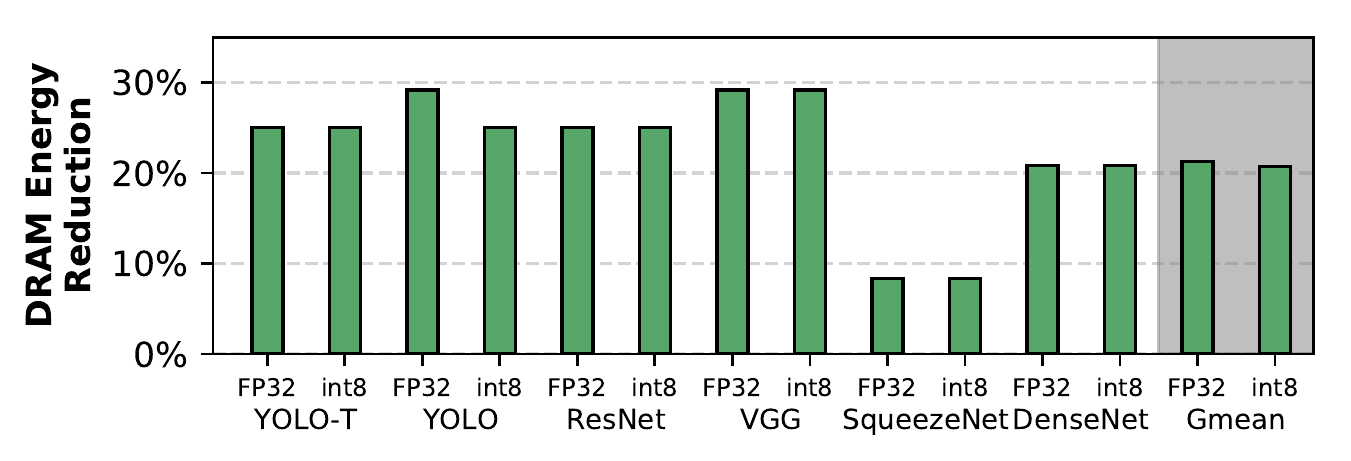}
\vspace{-0.8cm}
\caption{DRAM energy savings of \eden{}. We use FP32 and quantized int8 networks.}
\vspace{-0.2cm}
\label{fig:cpuenergy}
\end{figure}

\partitle{DRAM Energy} 
Figure~\ref{fig:cpuenergy} shows the DRAM energy savings of \eden{}, compared to a system with DRAM operating at nominal voltage and nominal latency. We make two observations. 
First, \eden{} achieves significant DRAM \energy{} savings across different DNN models. The average DRAM energy savings is 21\% across all workloads, and 29\% each for YOLO and VGG.
Second, the DRAM energy savings for FP32 and int8 are roughly the same, because the voltage {reduction} is very similar for both precisions (see Table~\ref{table:blanketresults}). 
 
We also perform evaluations for a target accuracy that is the same as the original. Our results show that EDEN enables {an average DRAM energy reduction} of 16\% (up to 18\%).

We conclude that EDEN is effective at saving DNN inference energy by reducing voltage while maintaining the DNN accuracy {within} 1\% of the original.

\begin{figure}[h]
\centering
\vspace*{-0.3cm}
\includegraphics[width=0.48\textwidth]{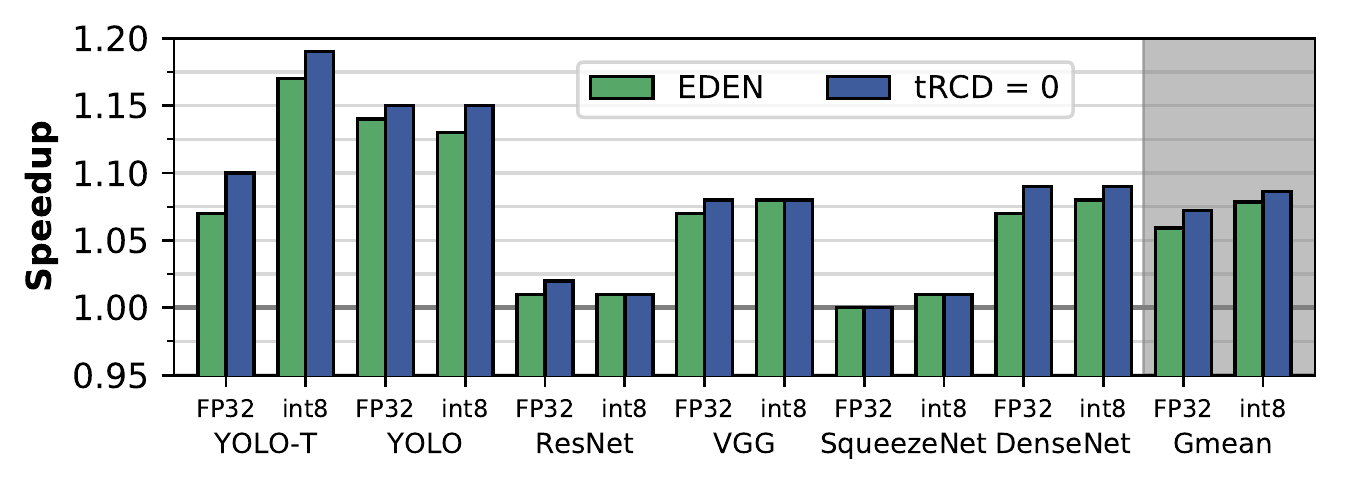}
\vspace{-0.8cm}
\caption{Speedup of \eden{} over baseline and versus a system with ideal activation latency. We use FP32 and an  quantized int8 networks.}
\vspace*{-0.3cm}
\label{fig:cpuspeedup}
\end{figure}

\partitle{Performance} Figure~\ref{fig:cpuspeedup} shows the speedup of \eden{} when we {reduce} $t_{RCD}$, and the speedup of a system with a DRAM module {that has} ideal $t_{RCD}=0$, compared to a system that uses DRAM with nominal timing parameters. We make three observations. First, YOLO DNNs exhibit high speedup with \eden{}, reaching up to 17\% speedup. The results of YOLO are better than the average because {YOLO} is more sensitive to DRAM latency. This is because some steps in YOLO {(e.g., Non-Maximum Suppression~\cite{non-maximum-suppression,non-maximum-suppression2}, confidence and IoU thresholding~\cite{yolo,rezatofighi2019generalized})} {perform} arbitrary indexing {into matrices} that lead to random memory {accesses,} which cannot easily be predicted by the prefetchers . Second, the average speedup of \eden{} (8\%) is very close to the average speedup of the ideal system with $t_{RCD}=0$ (10\%).  Third, we find that SqueezeNet1.1 and ResNet101 exhibit very little maximum theoretical speedup because they are not bottlenecked by memory latency.  

We also perform evaluations for a target accuracy that is the same as the original. Our results show that EDEN enables {an average performance gain} of 4\% (up to 7\%).

We conclude that \eden{} is effective at improving DNN inference performance by reducing DRAM latency while maintaining the DNN accuracy {within} 1\% of the original, especially on DNNs that are sensitive to memory latency.

\subsection{Accelerators}
We evaluate \eden{} on three different accelerators: GPU~\cite{oh2004gpu}, Eyeriss~\cite{eyeriss}, and TPU~\cite{tpu}.

\partitle{GPU Inference}
We evaluate EDEN on a GPU using the cycle-accurate {GPGPU-Sim} simulator \cite{gpgpusim}. We use GPUWattch~\cite{gpuwattch} to evaluate the {overall GPU} energy consumption. Table~\ref{tab:gpgpuconfig} details the NVIDIA Titan X GPU model~{\cite{titan}} we use in our evaluation. We use the reduced $t_{RCD}$ and $V_{DD}$ values that provide $<1\%$ accuracy degradation ({as} listed in Table~\ref{table:blanketresults}). We adapt four DarkNet-based binaries to run inference on the FP32/int8 YOLO and YOLO-Tiny DNNs.

 \begin{table}[h!]
 \small
   \vspace*{-0.1cm}
 \begin{tabular}{r|p{6cm}}
\textbf{Shader Core} & 28 SMs, 1417 MHz, 32 SIMT Width,\\
                     & 64 Warps per SM, 4 GTO Schedulers per Core\\
\textbf{Private L1 Cache} & 24 KB per SMM, Cache Block Size 128B  \\ 
\textbf{Shared Memory} &  96 KB, 32 Banks. Shared L2 Cache: 3MB \\ 
\textbf{Main Memory} & GDDR5, 2500MHz, 6 channels, 24 chips\\ 
 \end{tabular}
 \vspace{0.5em}
  \caption{Simulated {NVIDIA} Titan X GPU configuration
  }
 \label{tab:gpgpuconfig}
  \vspace*{-0.6cm}
 \end{table}

Our {results show that  EDEN provides  37\% average energy reduction (41.7\% for YOLO-Tiny, and 32.6\% for YOLO) compared to a GPU that uses DRAM with nominal parameters.}

Our {results also show that EDEN provides 2.7\%  average speedup  (5.5\% for the YOLO-Tiny, and 0\% for YOLO) compared to a GPU that uses DRAM with nominal parameters.}
DRAM with ideal $tRCD$ ($t_{RCD}=0$) provides  $6\%$ speedup for YOLO-Tiny and 2\% speedup for YOLO. These results indicate that 1) the YOLO DNN family is not DRAM latency bound {in our evaluation configuration}, and 2) \eden{} can achieve close to {the} ideal speedup of zero {activation} latency when the {DNN} is latency bound.

\partitle{Neural Network Inference Accelerators}
We evaluate \eden{} on Eyeriss~\cite{eyeriss} and Google's Tensor Processing Unit (TPU)~\cite{tpu} using the cycle-accurate SCALE-Sim simulator \cite{scalesim}. We use DRAMPower~\cite{drampower} to obtain DRAM \energy{} consumption from memory traces produced by SCALE-Sim. We use the built-in int8 AlexNet and YOLO-Tiny models and their accelerator-specific dataflows. We use DRAM parameters that yield a maximum accuracy loss of 1\% (Table~\ref{table:blanketresults}). Table~\ref{tab:eyerisstpuconfig} details the configuration of the Eyeriss and TPU inference accelerators. Eyeriss has an array of 12x14 processing elements (PEs) with {a} 324KB SRAM {buffer} for all data types (i.e., IFMs, weights and OFMs), and {the} TPU has an array of 256x256 PEs with {a} 24MB SRAM {buffer} for all data types. We evaluate both accelerators with DDR4 and LPDDR3 DRAM configurations, using Alexnet and YOLO-Tiny workloads.

 \begin{table}[h!]
    \centering
    \vspace{-0.15cm}
    \small
    \begin{tabular}{l|ll}
        
         & \textbf{Eyeriss} & \textbf{TPU} \\
        \hline 
        \textbf{Array} & $12\times14$~PEs & $256\times256$~PEs \\
        \textbf{{SRAM Buffers}}  & 324~KB & 24~MB\\
        \textbf{Main Memory}  & 4GB DDR4-2400 & 4GB DDR4-2400 \\
              &  4GB LPDDR3-1600 & 4GB LPDDR3-1600 \\
    \end{tabular}
    \vspace{0.5em}
    \caption{Simulated Eyeriss and TPU configuration\taha{s}.} 
    \label{tab:eyerisstpuconfig}
    \vspace{-0.7cm}
\end{table}

Our results show that  reducing the voltage level in DDR4 DRAM leads to significant DRAM \energy{} reductions on both Eyeriss and TPU accelerators.
\eden{} provides 1) 31\% average DRAM \energy{} savings {on} Eyeriss (31\% for YOLO-Tiny, and 32\% for Alexnet), and 2) 32\% average DRAM \energy{} savings {on} TPU (31\% for YOLO-Tiny, and 34\% for Alexnet).

Our results with {a} reduced voltage {level} in LPDDR3 are similar to those with DDR4. \eden{} provides an average DRAM \energy{} reduction of 21\%  for both Eyeriss and TPU accelerators running YOLO-Tiny and Alexnet. By using the accelerator/network/cache/DRAM \energy{} breakdown provided by {the} Eyeriss evaluations on AlexNet \cite{eyeriss_pub2}, we estimate that \eden{} can provide 26.8\% system-level \energy{} reduction on fully-connected layers and 7\% system-level \energy{} reduction on convolutional layers.

Our results with reduced {$t_{RCD}$} in LPDDR3 and DDR4 show that Eyeriss and TPU exhibit no speedup {from} reducing $t_{RCD}$. We observe that prefetchers are very effective in these architectures because the memory access patterns in the evaluated DNNs are {very predictable}.

\section{Related Work}
\label{sec:priorwork}

To our knowledge, this paper is the first to propose a general framework that reduces energy consumption and increases performance of DNN inference by using approximate DRAM with reduced voltage and latency. \eden{} introduces a new methodology to improve DNN's tolerance to approximate DRAM errors which is based on DNN error tolerance characterization and a new curricular retraining mechanism. We demonstrate the effectiveness of \eden{} by using error patterns that occur in real approximate DRAM devices.

In this section, we discuss closely related work on 1) approximate computing hardware for DNN workloads, and 2) modifying DRAM parameters.

\partitle{Approximate Computing Hardware for DNN Workloads}
\sloppy Many prior works propose to use approximate computing hardware for executing machine learning workloads~\cite{defect-tolerant,tnn95,ISSCC17,ISQED17,sram-aprox,retraining-based2015,approx2,approxnn1,approxnn2,storagemedia,unreliablememorycnn,faultresilient,resiliencertlnn,nnfaultmitigation,ares,understandingnnerrors1,reliabilitycnn,understandingnnerrors2}. All these works propose techniques for improving DNN tolerance for different types of approximate hardware mechanisms and error injection rates. Compared to these works, \eden{} is unique in 1) being the first work to use approximate DRAM with reduced voltage and latency, 2) being the first demonstration of DNN inference using error characterization of real approximate DRAM devices, 3) using a novel curricular retraining mechanism that is able to customize the DNN for tolerating \emph{high} error rates injected by the target approximate DRAM, and 4) mapping each DNN data type to a DRAM partition based on the error tolerance of the DNN data type and the bit error rate of the DRAM partition. We classify related works on approximate {hardware} for DNN workloads into six categories.

First, works that reduce DRAM refresh to save DNN energy~\cite{rana,stdrc,approxnnrefresh}. RANA~\cite{rana} and St-DRC~\cite{stdrc} propose to reduce DRAM refresh rate in the embedded DRAM (eDRAM) memory of DNN accelerators. Nguyen et al. \cite{approxnnrefresh} propose to apply similar refresh optimization techniques to off-chip DRAM in DNN accelerators. These mechanisms use customized retraining mechanisms to improve the accuracy of the DNN in the presence of a moderate amount of errors.

Second, works that study the error tolerance of neural networks to uniform random faults in SRAM memory~{\cite{ares,understandingnnerrors1,reliabilitycnn,understandingnnerrors2,Salami:2018}}. For example, Li et al.~\cite{understandingnnerrors2} analyze the effect of various numeric representations on error tolerance. Minerva \cite{minerva} proposes an algorithm-aware fault mitigation technique to mitigate the effects of low-voltage SRAM in DNN accelerators. \loiscomment{Improve, this description is bad}

Third, works that study approximate arithmetic logic in DNN workloads~\cite{thundervolt,resiliencertlnn,resiliencertlnn,nnfaultmitigation}. ThUnderVolt~\cite{thundervolt} proposes to underscale the voltage of arithmetic elements. Salami et al.~\cite{resiliencertlnn} and Zhang et al. \cite{nnfaultmitigation} present fault-mitigation techniques for neural networks that minimize errors in faulty registers and logic blocks with pruning and retraining.\loiscomment{improve this}

Fourth, works that study approximate emerging memory technologies for neural network acceleration. Panda et al. \cite{approxneuromorphic1} and Kim \cite{approxneuromorphic2} propose neuromorphic accelerators that use spintronics and memristors to run a proof-of-concept fuzzy neural network. 

Fifth, works that study the effects of approximate storage devices on DNN workloads~\cite{storagemedia,faultresilient}. Qin et al.~\cite{storagemedia} study the error tolerance of neural networks that are stored in approximate {non-volatile} memory (NVM) media. The authors study the effects of turning the ECC off in {parts of} the NVM media that {store} the neural network data. Wen et al. \cite{faultresilient} propose to mitigate the effects of unreliable disk reads with a specialized ECC variant that aims to mitigate error patterns present in weights of shallow neural networks.

\loiscomment{Check this paragraph. Are these works correctly classified?}
Sixth, works that study the intrinsic error resilience  of DNNs by injecting   {randomly-distributed} errors in DNN data~\cite{defect-tolerant,unreliablememorycnn, unreliablememorycnn,approxnn1,approxnn2,resiliencertlnn,nnfaultmitigation}. These works assume that the errors can come from any component of the system (i.e., they do not target a specific approximate hardware component). Marques et al. \cite{unreliablememorycnn} study the accuracy of DNNs under different error injection rates and {propose} various error mitigation techniques. This work uses a simple probabilistic method to artificially inject errors into the DNN model.
ApproxANN~\cite{approxnn1}  uses an algorithm that optimizes the DNN accuracy by taking into account the error tolerance and the {criticality} of each component of the network. The quality-configurable Neuromorphic Processing Engine (qcNPE)~\cite{approxnn2} uses processing elements with dynamically configurable accuracy for executing approximate neural networks.

\partitle{Modifying DRAM Parameters}
Many prior works study the effects of modifying DRAM parameters on reliability, performance and energy consumption. We already discuss some prior works that reduce DRAM voltage, access latency, and refresh rate in Section~\ref{sec:background_reducing}. EDEN leverages the characterization techniques introduced in Voltron~\cite{voltron} and Flexible-Latency DRAM~\cite{flydram} to perform the DRAM characterization required to map a DNN to approximate DRAM with reduced voltage and reduced latency (Section~\ref{sec:netmapping}). We classify other related works that modify DRAM parameters into three categories.

First, works that aim to characterize and reduce energy consumption at reduced supply voltage levels~\cite{david2011memory,freqscaling2,voltron,vampire}. {David et al.}~\cite{david2011memory} {propose} memory dynamic voltage and frequency scaling (DVFS) {to reduce} DRAM power. MemScale \cite{freqscaling2} {provides} dynamic voltage and/or frequency scaling {in} {main} memory {to reduce} energy consumption, while meeting a maximum tolerable performance degradation. 
Voltron~\cite{voltron} studies {voltage reduction in} real {DRAM} devices in detail and proposes solutions to reduce voltage {reliably} based on {observed error characteristics and system performance requirements}.
VAMPIRE \cite{vampire} proposes a new DRAM power model that is based on {the characteristics of} real DRAM devices.

Second, works that investigate DRAM characteristics under reduced access latency ~\cite{flydram, voltron, solardram, diva, aldram, goossens,drange,puf}. Adaptive-Latency DRAM \cite{aldram} characterizes the guardbands present in timing parameters defined by DRAM manufacturers, and exploits the extra  timing margins to {reliably} reduce DRAM latency across different chips and temperatures. Flexible-Latency DRAM \cite{flydram} analyzes the spatial distribution of reduced-latency-induced cell failures, and uses this information to {reliably} access different regions of DRAM with different timing parameters. {DIVA-DRAM \cite{diva}} proposes an automatic method for finding the lowest reliable operation latency of DRAM, via a combination of runtime profiling and ECC.

Third, works that aim to reduce DRAM latency by modifying the microarchitecture of DRAM or the memory controller~{\cite{hassan2019crow, tldram, wang2018reducing, chargecache,lu2015improving,Son:2013,Choi:2015,decoupled-dma}}. These works reduce latency without introducing bit errors.
\section{Conclusion}
\label{sec:conclusion}
This paper introduces EDEN, the first general framework that enables energy-efficient and high-performance DNN inference via approximate DRAM, while strictly meeting a target DNN accuracy. \eden{} uses an iterative mechanism that profiles the DNN and the target approximate DRAM with reduced voltage and timing parameters. \eden{} improves DNN accuracy with a novel curricular retraining mechanism that tolerates high bit error rates. We evaluate \eden{} in both simulation and on real hardware. Our evaluation shows that \eden{} enables 1) an average DRAM energy reduction of {21\%, 37\%}, 31\%, and 32\% in CPU, GPU, Eyeriss, and TPU architectures, respectively, {across a variety of {state-of-the-art} DNNs,} and 2)  \lois{average (maximum)} performance {gains} of 8\% \lois{(17\%)} in {CPUs} and 2.7\% \lois{(5.5\%)} in {GPUs,} for latency-bound DNNs. We expect that the core principles of EDEN generalize well across different memory devices, memory parameters, and memory technologies. We hope that EDEN enables further research {and development} on the use of approximate memory for machine learning workloads.

\section*{Acknowledgments}
We thank the anonymous reviewers for feedback. We thank the SAFARI Research Group members for feedback and the stimulating intellectual environment they provide. We acknowledge the generous funding provided by our industrial partners: Alibaba, Facebook, Google, Huawei, Intel, Microsoft, and VMware. This research was supported in part by the Semiconductor Research Corporation. Skanda Koppula was supported by the Fulbright/Swiss Government Excellent Scholarship 2018.0529.

\bibliographystyle{IEEEtranS.bst}
\balance
\bibliography{refs}

\end{document}